\documentclass[apj]{emulateapj}
\usepackage{apjfonts}

\gdef\date{August 11}
\slugcomment{accepted for publication, ApJ, 2008, 689}

\shorttitle{Surface density dependent evolution to $z=3$}

\shortauthors{Franx et al.}

\begin{document}


\title{Structure and star formation in galaxies out to $z=3$: evidence
  for surface density dependent evolution and
  upsizing\altaffilmark{1,2}}


\author{
Marijn   Franx\altaffilmark{3},
Pieter~G.~van Dokkum\altaffilmark{4},
Natascha M. F\"orster Schreiber\altaffilmark{5},
Stijn Wuyts\altaffilmark{6,7},
Ivo Labb\'e\altaffilmark{8,9},
Sune Toft\altaffilmark{10}
}

\altaffiltext{1}
{Based on observations with the NASA/ESA {\em Hubble Space
Telescope}, obtained at the Space Telescope Science Institute, which
is operated by AURA, Inc., under NASA contract NAS 5--26555.}
\altaffiltext{2}
{Based on observations collected at the European Southern Observatory,
  Chile (ESO Programme LP168.A-0485)}
\altaffiltext{3}
{Leiden Observatory, Leiden University, P.O. Box 9513, NL-2300 RA Leiden,
  Netherlands}
\altaffiltext{4}
{Department of Astronomy, Yale University, New Haven, CT 06520-8101}
\altaffiltext{5}{MPE, Giessenbackstrasse, D-85748, Garching, Germany}
\altaffiltext{6}{Harvard-Smithsonian Center for Astrophysics, 60
  Garden Street,
Cambridge, MA 02138}
\altaffiltext{7}{W. M. Keck postdoctoral fellow}
\altaffiltext{8}{Carnegie Observatories, 813 Santa Barbara Street,
  Pasadena, CA  91101}
\altaffiltext{9}{Hubble Fellow}
\altaffiltext{10}
{European Southern Observatory, D-85748 Garching, Germany}








\begin{abstract}
\def\car{{\hbox{}\par\parindent=0pt}}
\def\Msun{M_{\odot}}
\def\Mstar{M_*}
\def\re{r_e}
\def\mre{\Mstar/\re}
\def\omit#1{}
\def\yrinv{yr$^{-1}$}
\def\msunkpc{$\Msun$ kpc$^{-2}$}

We present an analysis of galaxies in the CDF-South.
We find a tight relation to $z=3$ between color and size at a given mass,
with red galaxies being small, and blue galaxies being large.
We show that the relation is driven by stellar surface density or
inferred velocity
dispersion:
galaxies with high surface density
are red and have low specific star formation rates, and galaxies with
low surface density are blue and have high
specific star formation rates. 
Surface density and inferred velocity dispersion are better correlated with
specific star formation  rate and color than stellar mass.
Hence stellar mass by itself is not a good predictor of the star 
formation history of galaxies.
In general, galaxies at a given
surface density have higher specific star formation rates at higher
redshift. 
Specifically, galaxies with a surface density of $1-3\ 10^9$ $M_\odot$ $kpc^{-2}$
are ``red and dead'' at low redshift,
approximately 50\% are forming stars at $z=1$, and almost all are
forming stars by $z=2$.
This provides direct additional evidence for the late evolution of
galaxies onto the red sequence.
The sizes of galaxies at a given mass evolve like $1/(1+z)^{0.59
\pm 0.10}$.
Hence galaxies undergo significant upsizing in
their history. The size evolution is fastest for the highest mass
galaxies, and
quiescent galaxies.
The persistence of the structural relations from $z=0$ to $z=2.5$, and
the upsizing of galaxies
imply that a relation analogous to the Hubble
sequence exists out to $z=2.5$, and possibly beyond.
The star forming galaxies at $z\ge 1.5$ are quite different from star
forming
galaxies at $z=0$, as they have likely very high gas fractions, and
star formation time scales comparable to the orbital time.

\end{abstract}





\section{Introduction}

\def\car{{\hbox{}\par\parindent=0pt}}
\def\Msun{M_{\odot}}
\def\Mstar{M_*}
\def\re{r_e}
\def\mre{\Mstar/\re}
\def\omit#1{}
\def\yrinv{yr$^{-1}$}
\def\msunkpc{$\Msun$ kpc$^{-2}$}

\def\figprezero{\figzero}
\def\figpreone{\figone}
\def\figpretwo{\figtwo}
\def\figprethree{\figthree}
\def\figprefour{\figfour}
\def\figprefive{\figfive}
\def\figpresix{\figsix}
\def\figpreseven{\figseven}
\def\figpreeight{\figeight}
\def\figprenine{\fignine}
\def\figpreten{\figten}
\def\figpreeleven{\figeleven}
\def\figpretwelve{\figtwelve}
\def\figpreAone{\figAone}

\def\figsubzero{}
\def\figsubone{}
\def\figsubtwo{}
\def\figsubthree{}
\def\figsubfour{}
\def\figsubfive{}
\def\figsubsix{}
\def\figsubseven{}
\def\figsubeight{}
\def\figsubnine{}
\def\figsubten{}
\def\figsubeleven{}
\def\figsubtwelve{}
\def\figsubAone{}

\def\figone{
\begin{figure*}[t]
\plotone{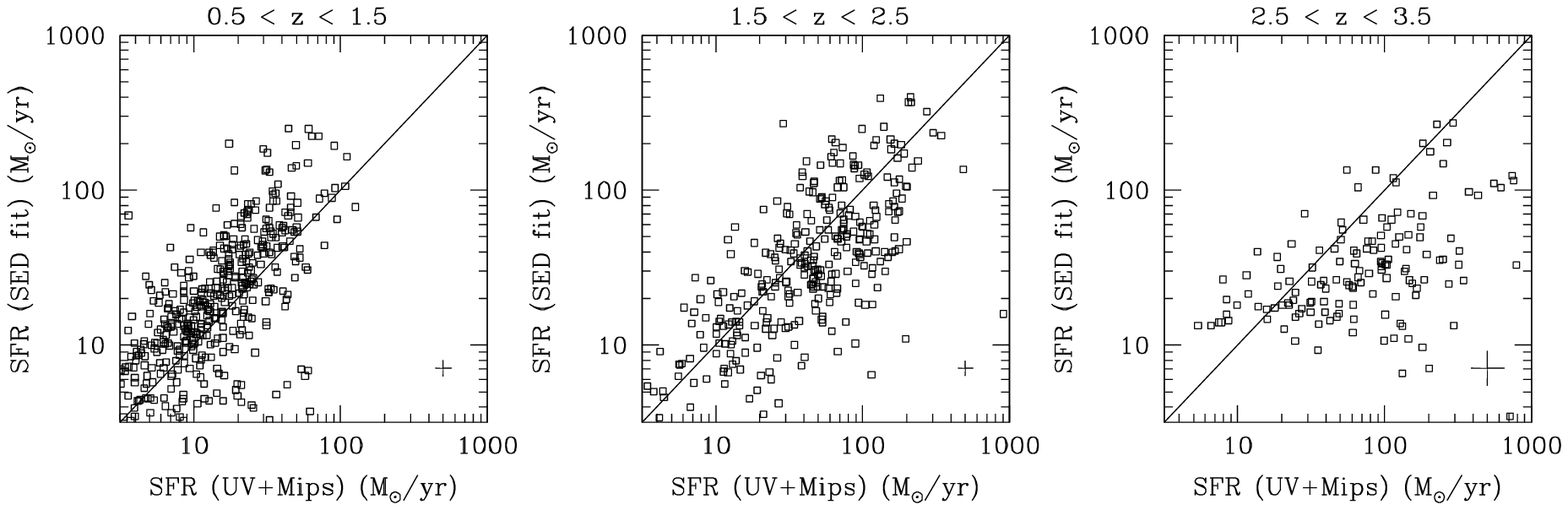}
\caption{\small
Comparison of the star formation rates estimated from the
UV + 24$\mu$m MIPS fluxes, and the star formation rates estimated by
SED fits. There is generally a good correspondence at redshifts
below 2.5, with mild offsets smaller than 30\%.
For the remainder of the paper, we use the UV + 24$\mu$m MIPS star
formation rates.
The typical formal error in the estimated star formation rates is shown
in the lower right corner.
The systematic uncertainties are a factor of 2 or more.
\label{figone}}
\end{figure*}
}

\gdef\figurenum#1{}
\def\figzero{
\begin{figure}[t]
\epsscale{0.9}
\plotone{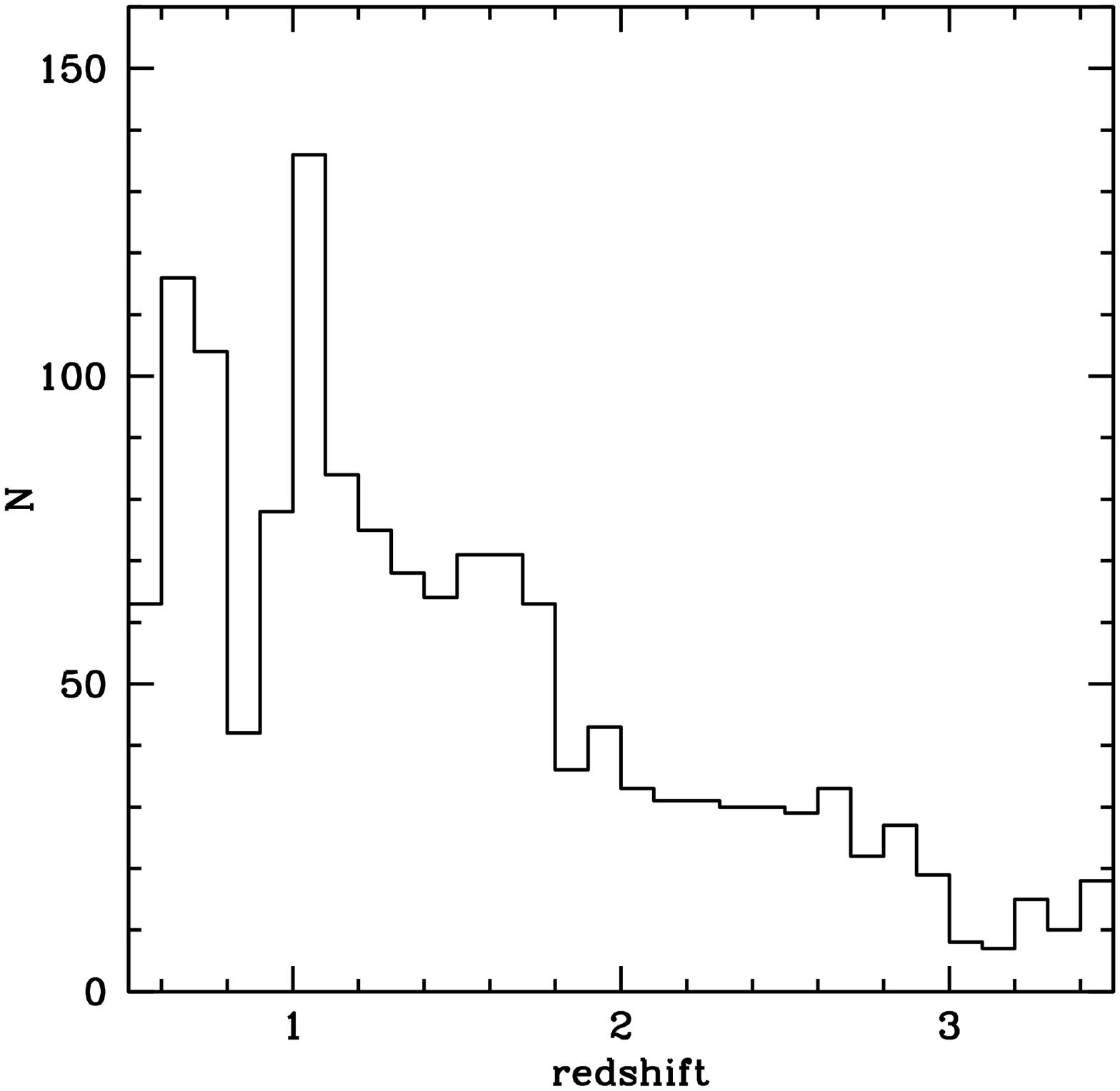}
\epsscale{1.0}
\caption{\small
The redshift distribution of the final CDFS sample. The
galaxies are  selected in the observed $K$ band, which lies redward of the
rest-frame Balmer-4000\AA\ break for this sample.
\label{figone}}
\end{figure}
}

\def\figtwo{
\begin{figure*}[t]
\epsscale{1.2}
\plotone{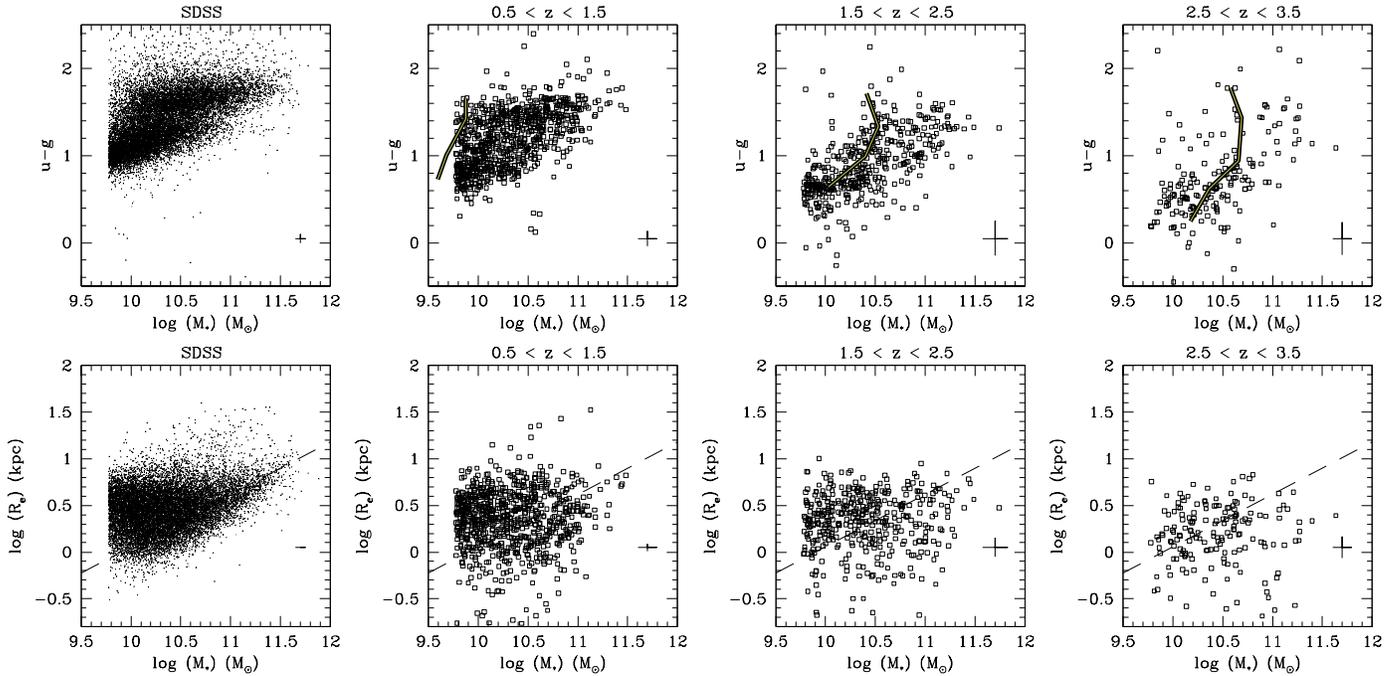}
\epsscale{1.0}
\caption{\small
Top row: the restframe $u-g$ color against stellar mass, for the low redshift
sample (from the SDSS), and the redshifts 1, 2 and 3.
A large spread in restframe $u-g$ colors exists for masses between
$10^{10}$ and $10^{11}$, for all redshifts.
The absence of red, low mass galaxies at $z\ge 2$ is due to selection
effects.
The thick curves in the higher redshift panels delineate the  area
where the samples are less than 75 \% complete.
The red galaxies to the upper left of the curve  are missing as they are too faint
in the observed near-IR.
Bottom row:
The relation between  size  and  stellar mass for our
redshift intervals.
A large spread in sizes exists for masses between
$10^{10}$ and $10^{11}$, for all redshifts.
The thin dashed line indicates the mass-size relation from Shen et
al. (2003) for early-type galaxies at $z=0$.
It is clear from the figure that the scatter in color and size is very
large
for masses between 10$^{10}$ and 10$^{11}$ $\Msun$.
As in subsequent plots, the  typical formal error is shown in the lower
right corner.
\label{figone}}
\end{figure*}
}

\def\figthree{
\begin{figure*}[t]
\epsscale{1.2}
\plotone{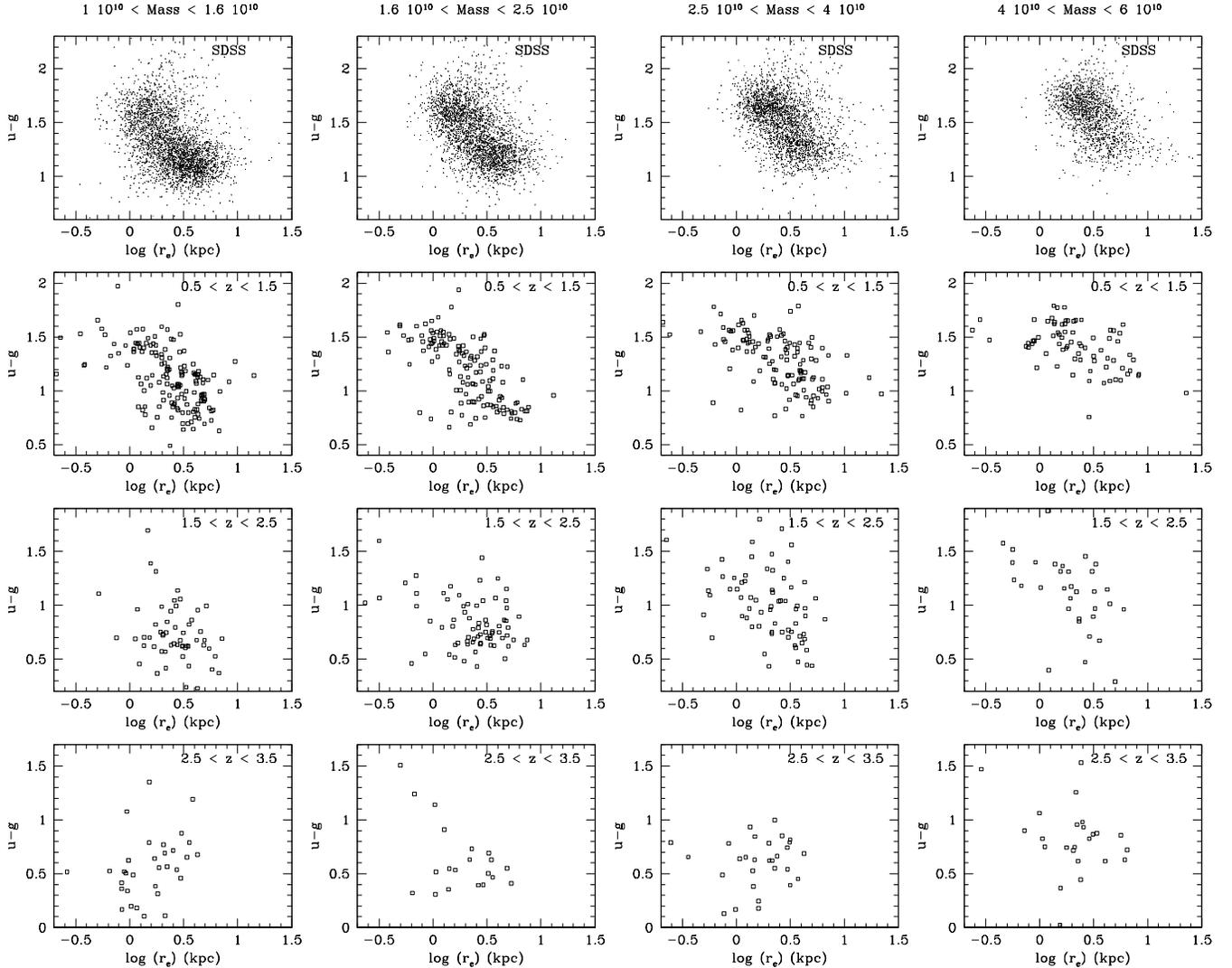}
\epsscale{1.00}
\caption{\small
The correlation between color and size, at fixed stellar
mass, for our sample. The top row shows the relation for the
SDSS sample, the row below for galaxies with $0.5<z < 1.5$, the 3rd
row for galaxies with $1.5< z < 2.5$, the bottom row for galaxies
with $2.5 < z < 3.5$.
The top rows show a strong correlation between color and size
in the mass bins,
with large galaxies being blue, and small galaxies being red.
This demonstrates that the scatter in color and size at a given
mass is driven by a third parameter.
A correlation is still present at $z\approx 2$, especially
in the mass bin $2.5\ 10^{10} \Msun < M_* < 4\ 10^{10}$.
At high redshift ($z>2.5$) the sample is too small to establish
whether or not a relation exists.
\label{figone}}
\end{figure*}
}

\def\figfour{
\begin{figure*}[t]
\figurenum{4a}
\epsscale{1.2}
\plotone{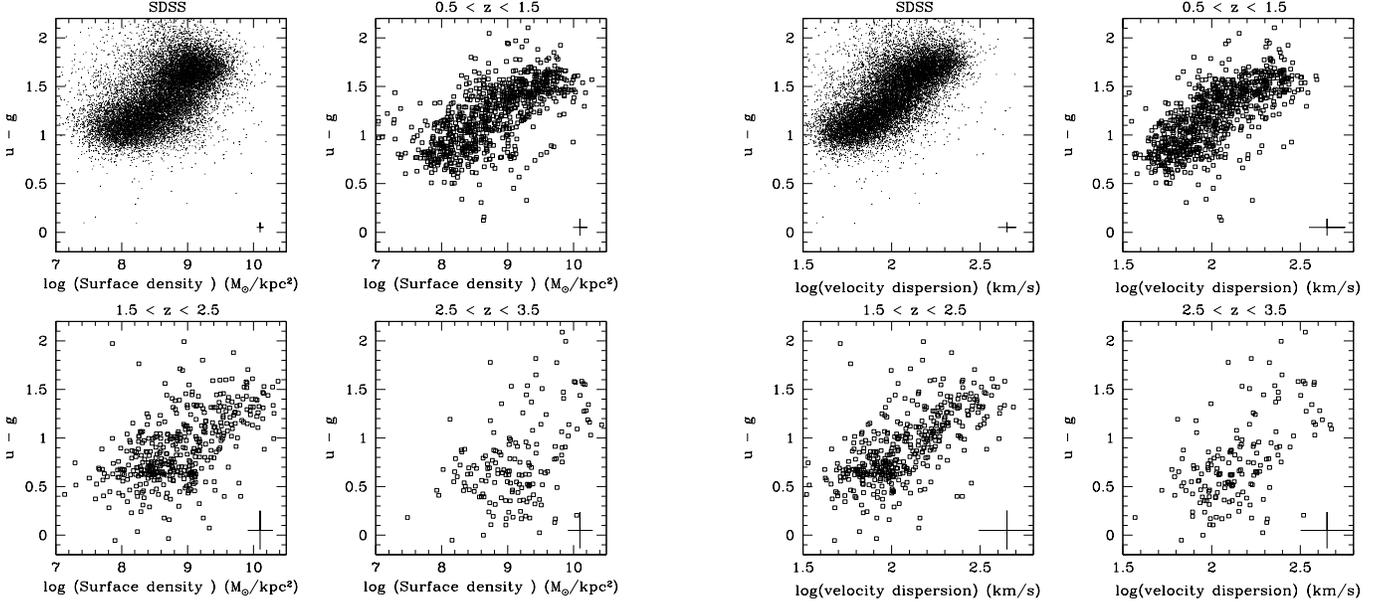}
\epsscale{1.0}
\caption{\small
Left panels: the correlation between color and surface density, at redshifts 0,
1, 2 and 3. At all redshifts, a correlation is found,
with high surface density galaxies being red, and low surface density galaxies
being blue.
This indicates that
surface density is one of the main driving parameters for galaxy evolution.
The color is more strongly correlated with surface density, than with
mass; and this holds out to the highest redshift.
Right panels: the correlation between color and velocity dispersion,
at the same redshifts. Again, a good correlation is found, with high velocity
dispersion galaxies being red, and low velocity dispersion galaxies being blue.
The velocity dispersions have been estimated from $M_*/r_e$.
\label{figone}}
\end{figure*}
}

\def\figfive{
\begin{figure*}[t]
\figurenum{5a}
\epsscale{1.2}
\plotone{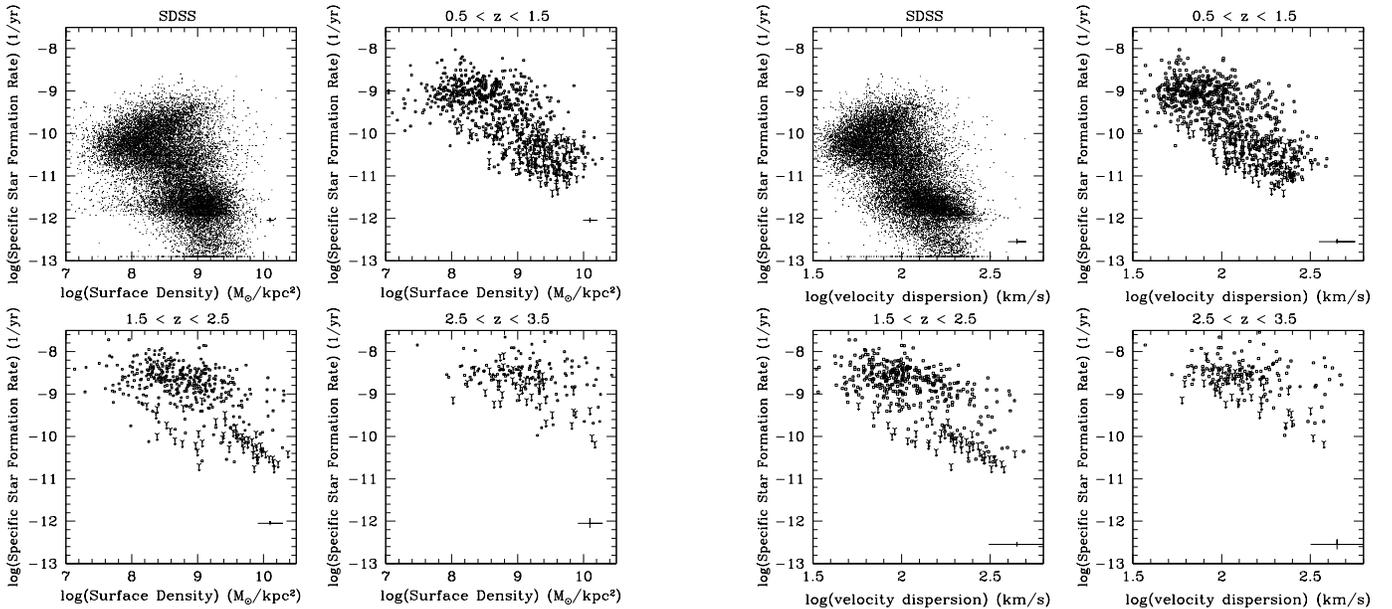}
\epsscale{1.0}
\caption{\small Left panels: the correlation between specific star formation rate and
surface density, at redshifts 0, 1, 2, and 3.
The SDSS specific star formation rate is based on an analysis of emission
lines, whereas the high redshift points are based on the 24 $\mu$m emission.
The arrows denote upper limits.
As can be seen, the highest surface density galaxies have low specific
star formation rates out to redshift of 2.5, showing that specific star formation rate is
causing to a large degree the relation between color and surface density.
The limiting specific star formation rates for the redshift of 3 galaxies are too
high to allow the detection of a correlation.
Right panels:
The correlation between velocity dispersion and
specific star formation rate. Again, a good correlation persists to
$z=2.5$, and above that the specific star formation rates are not
well enough determined.
\label{figone}}
\end{figure*}
}

\def\figsix{
\begin{figure*}[t]
\figurenum{6}
\epsscale{1.2}
\plotone{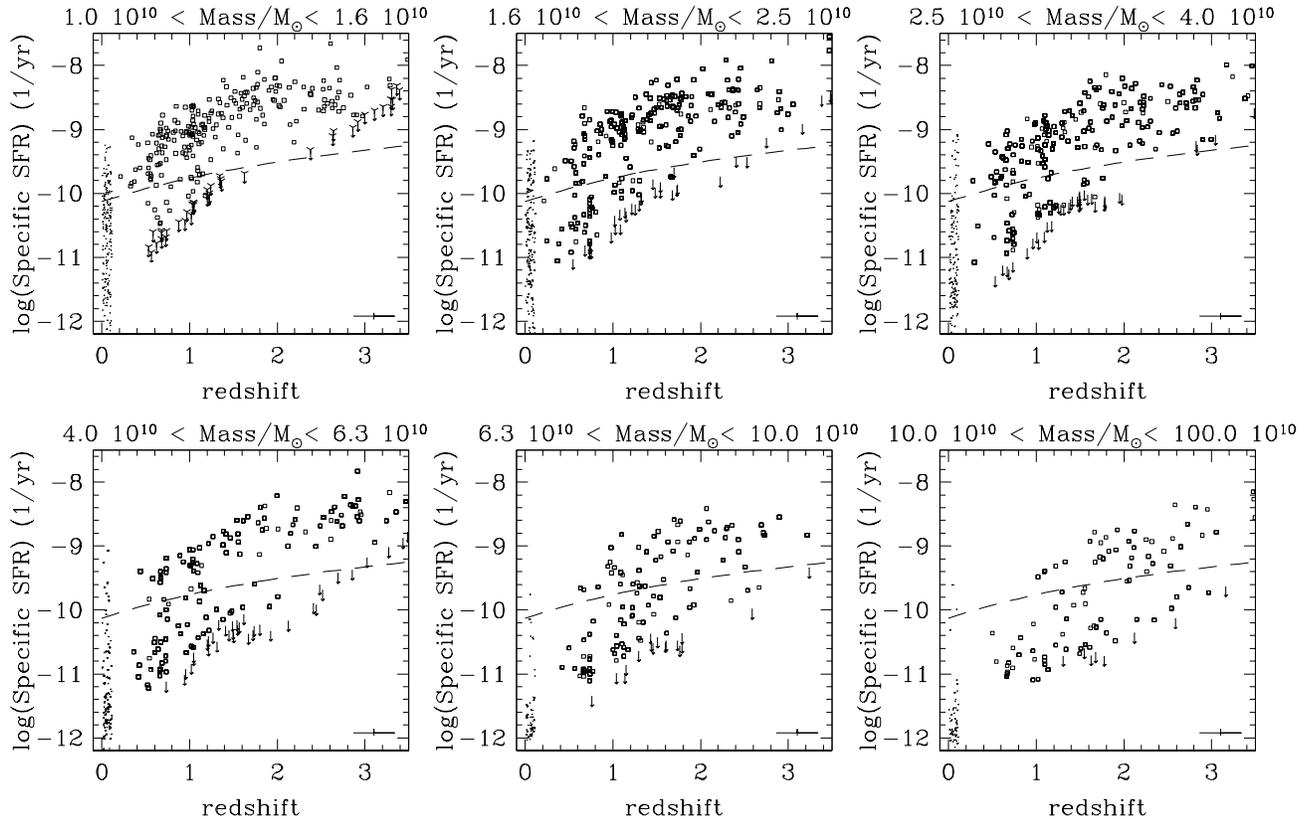}
\epsscale{1.0}
\caption{\small The evolution of the specific star formation rate in mass bins.
In all mass bins, the specific star formation rates increase with redshift,
and  the spread remains large.
The dashed line indicates the specific
star formation rate equal to $t_{hubble}^{-1}$.
\label{figone}}
\end{figure*}
}

\def\figseven{
\begin{figure*}[t]
\figurenum{7}
\epsscale{1.2}
\plotone{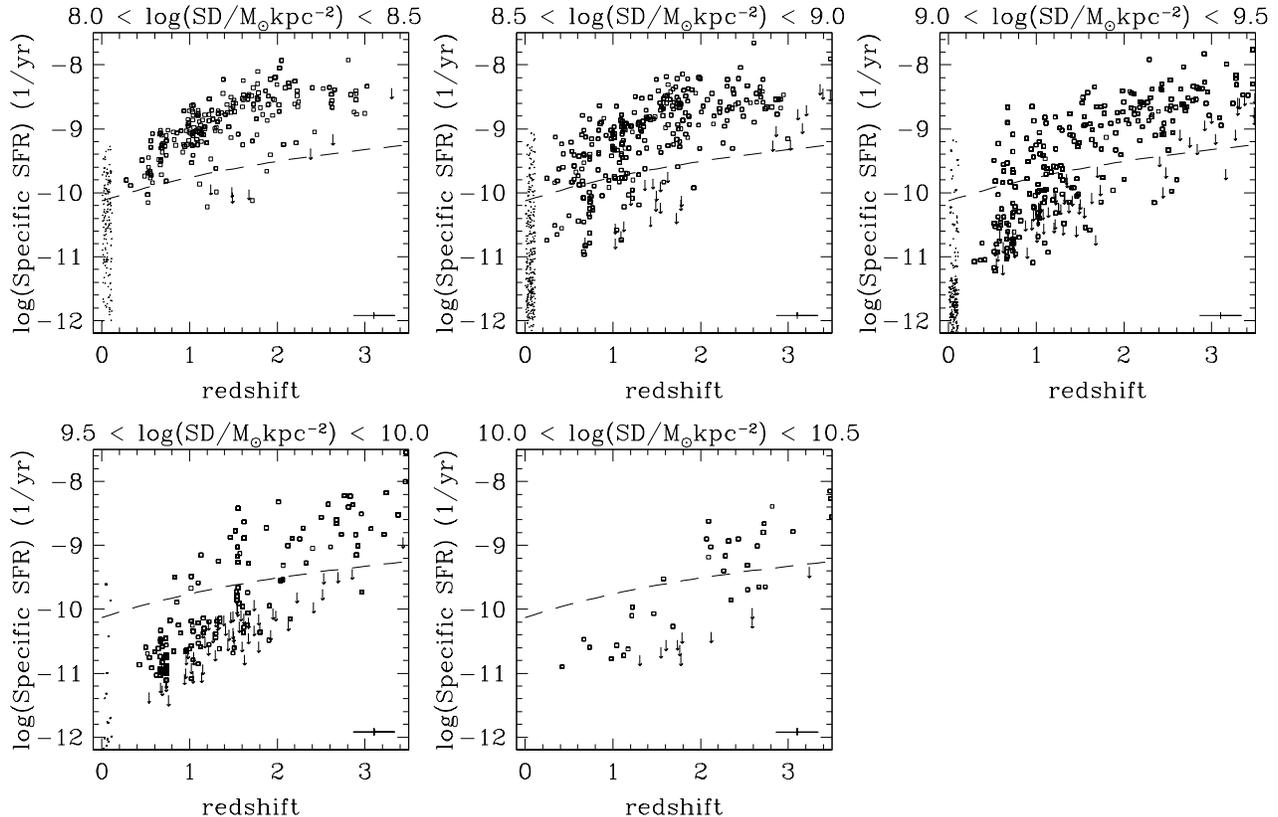}
\epsscale{1.0}
\caption{ \small The evolution of the specific star formation rate in bins of
surface density. The specific star formation rates increase in all
bins.
In the lowest surface density bin, the specific star formation rates
are always high, and very few quiescent galaxies exist.
We define quiescent galaxies as those having specific star formation
rates
below $0.3 \times t_{hubble}^{-1}$.
The dashed line indicates a specific star formation rate of $t_{hubble}^{-1}$.
In the intermediate surface density
bin with $ 10^9 < SD < 10^{9.5}$ the specific star formation rates are
very low at $z<0.2$, increase rapidly to high values at $1=0.8-1.5$,
to have only high values at $z \ge 1.5$.
The highest surface density bins always have the lowest specific star formation
rates.
\label{figone}}
\end{figure*}
}

\def\figeight{
\begin{figure*}[t]
\figurenum{8}
\epsscale{1.2}
\plotone{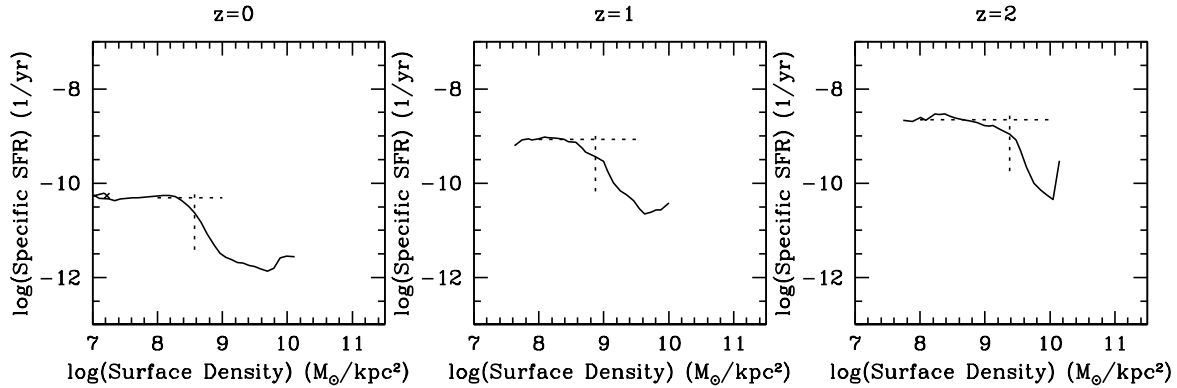}
\epsscale{1.0}
\caption{\small The median specific star formation rate as a function of
stellar surface density, at 3 redshift intervals: local galaxies from SDSS,
and galaxies at $0.5<z<1.5$, and $1.5<z<2.5$.
The median specific star formation rate is fairly constant at low
surface densities, to decrease rapidly at higher surface densities.
We have defined the threshold density, where the specific star formation
rate is a factor of 3 lower than the specific star formation rate of the
plateau.
\label{figone}}
\end{figure*}
}

\def\fignine{
\begin{figure*}[t]
\figurenum{9}
\epsscale{1.2}
\plotone{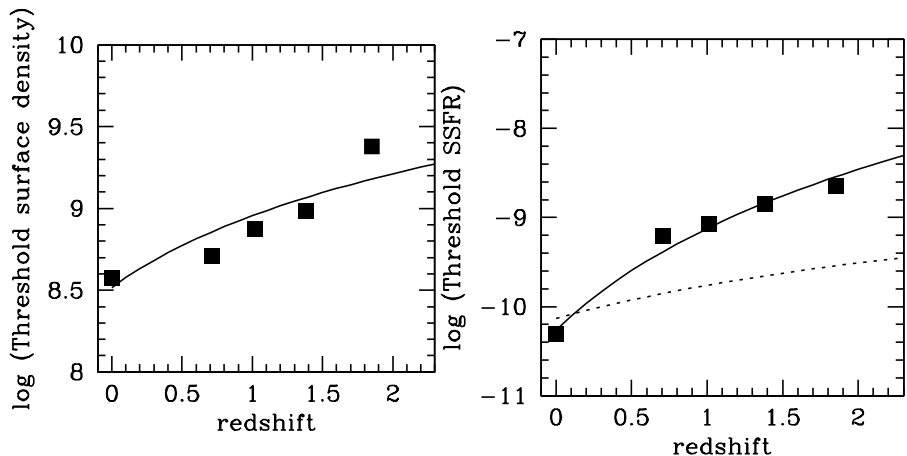}
\epsscale{1.0}
\caption{\small a) The evolution of surface density threshold, above which the
specific star formation rate is low. At higher redshifts, the threshold
has been measured in bins of width $\delta z = 1$.
The threshold evolves fast, close to $(1+z)^{1.5}$, as shown by the curve.
b). The evolution of the specific star formation rate below the surface
density threshold. The specific star formation rate evolves very rapidly
like $(1+z)^{3.8}$, as shown by the curve.
The dotted line curve the specific star formation rate equal to
$t_{hubble}^{-1}$.
\label{figone}}
\end{figure*}
}

\def\figten{
\begin{figure*}[t]
\figurenum{10}
\epsscale{1.2}
\plotone{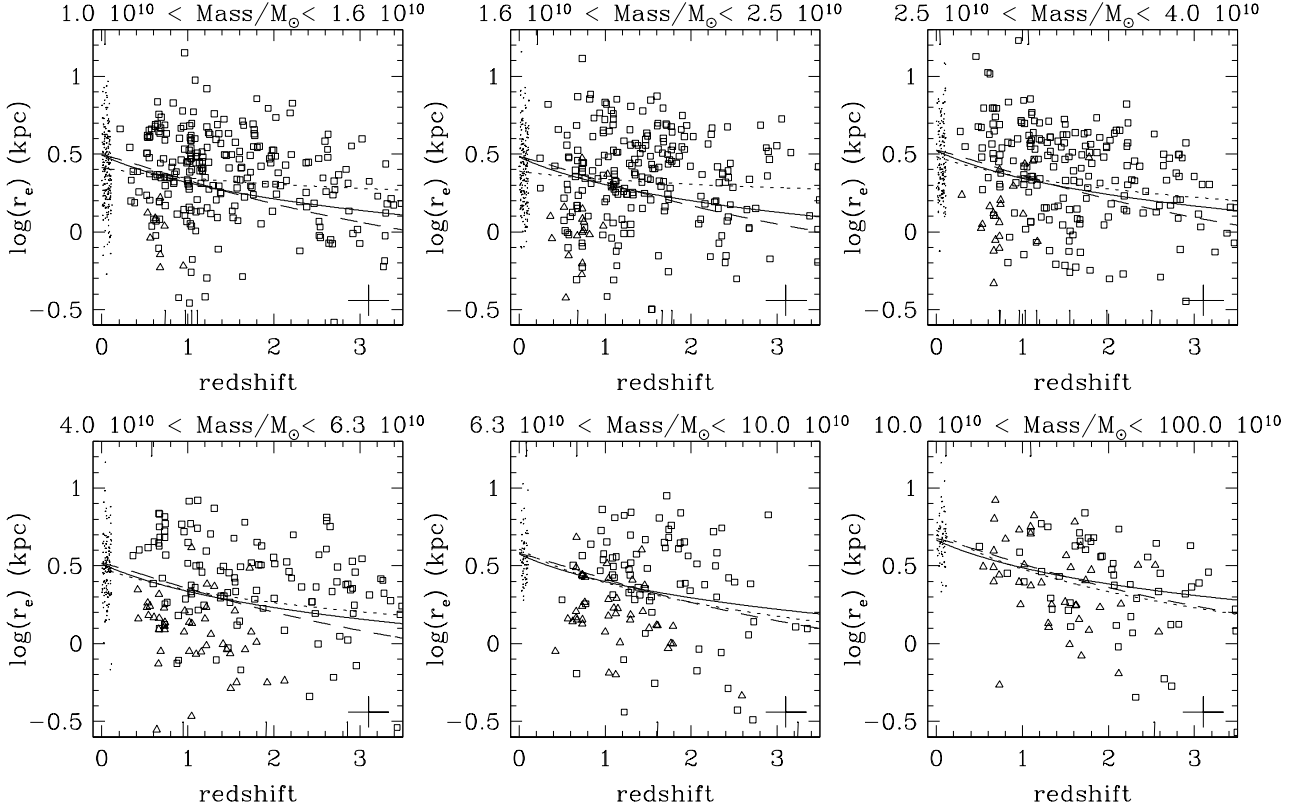}
\epsscale{1.0}
\caption{\small The evolution of half-light radius with redshift, in narrow mass bins.
The sizes steadily decreases with  redshift, both for high mass and low
mass galaxies. The curves show fits using the average coefficient
 $r_e \propto (1+z)^{-0.59}$. The dashed curves show the fit using
$r_e \propto H(z)^{-2/3}$, which is predicted from simple formation
models.
The dotted curves show the fits of $\re \propto 1/(1+z)^\alpha$ to the
individual bins. The index $\alpha$ varies from 0.14$\pm$0.07 to
0.81$\pm$ 0.1, from the
low mass to the high mass bin. The average index is 0.59 $\pm$ 0.10.
The triangles are quiescent galaxies with low specific star formation rates
(as defined in the text). They evolve faster, like $\re \propto
1/(1+z)^{1.22\pm 0.15}$.
Overall, the sizes evolve significantly with redshift, the strongest
for the high mass galaxies, and the quiescent galaxies.
\label{figone}}
\end{figure*}
}

\def\figeleven{
\begin{figure*}[t]
\figurenum{11}
\epsscale{1.0}
\plotone{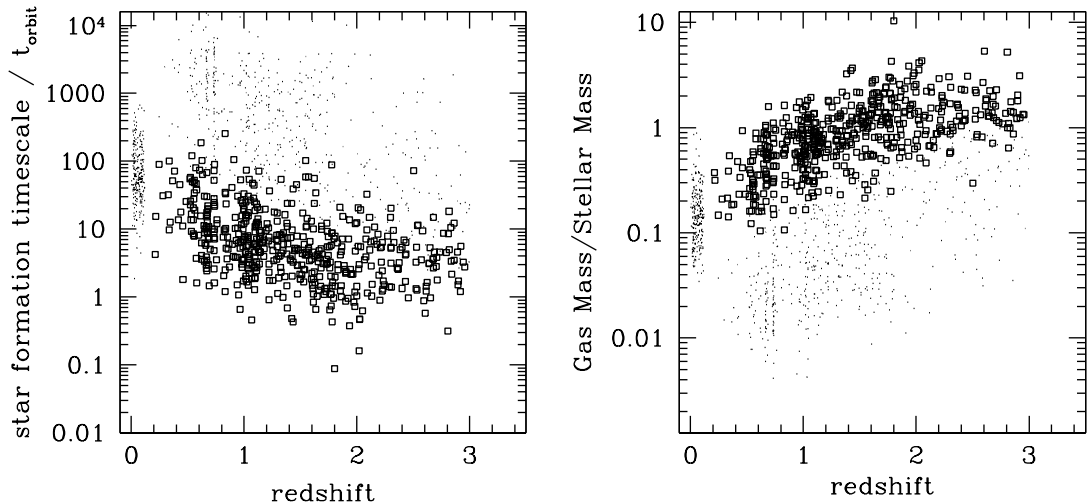}
\epsscale{1.0}
\caption{\small
Left panel: The ratio of star formation time scale to dynamical time scale.
Whereas at low redshift, the dynamical timescale is generally more
than 30 times
longer than the star formation time scale , it decreases to values of
3-10 at a redshift of 1.5 and above for the strongly star forming galaxies.
For these galaxies, the gas has barely time to settle in a disk.
Right panel: The evolution of the ratio of gas mass to stellar mass
 with redshift.
The gas mass has been estimated from the star formation rate, and
the kennicutt star formation law.
Galaxies with significant star formation are indicated with a large symbol
(specific star formation $>$ 1/3 threshold specific star formation rate).
Whereas at low redshift the typical gas fraction is around 10\%, it
increases steadily with redshift, to reach values of 0.3-1 at $z=1.5$
and above. The most strongly star forming galaxies at $z=1.5$ and
above have gas masses comparable to their stellar masses.
\label{figone}}
\end{figure*}
}

\def\figtwelve{
\begin{figure*}[t]
\figurenum{12}
\epsscale{1.0}
\plotone{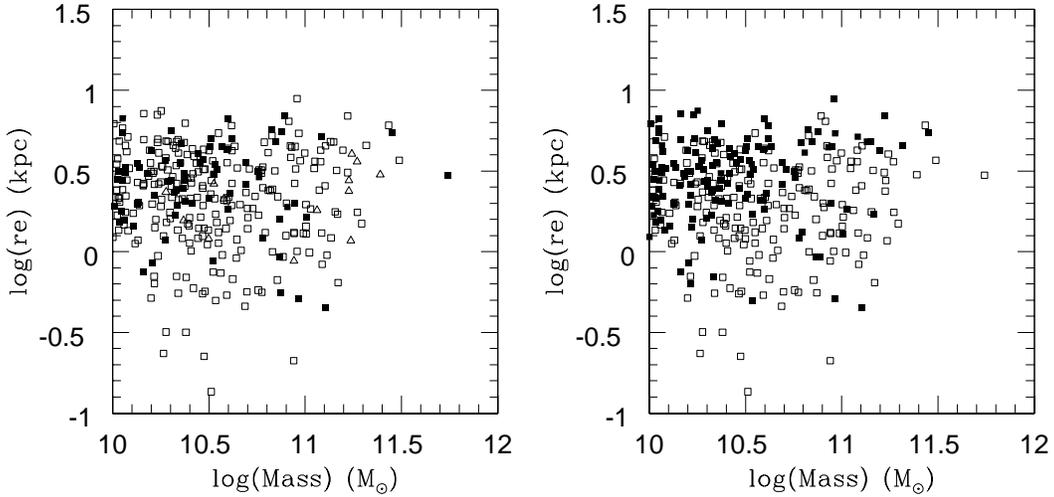}
\epsscale{1.0}
\caption{\small
Simulated selection effects for our $z=2$ sample.
The left panel shows the selection on  observed $H_\alpha$ flux.
The $H_\alpha$ flux has been estimated from the
star formation rate, and extinction corrected, assuming the extinction
from the SED fit.
Galaxies with H$\alpha$ flux $>$ $10^{42}$ ergs sec$^{-1}$  and $V_{ab} < 25.5$
are indicated  with full squares, the galaxies with
H$\alpha$ flux $>$ $10^{42}$ ergs sec$^{-1}$ and $V_{ab} > 25.5$ are
indicated with open triangles, and the other galaxies are indicated
with open squares. The V band limit is used as most often Near-IR
spectroscopy is done on galaxies with optical redshifts measured
first.
The right panel shows the selection on  UV-slope and
UV continuum flux
(right), simular to those used for Ly-break and BM-BX galaxies
(e.g., Steidel et al. 2006).
Filled squares are galaxies with $V_{ab} < 25.5 $ and $B-V < 1.2$.
In both panels, we see that the selected galaxies are prefentially
large
compared to the sample as a whole. It is obviously difficult to study
galaxies at the full range of sizes, given the mass.
We note that our full sample is also likely to be incomplete at low
masses
and small sizes.
\label{figone}}
\end{figure*}
}

\def\figAone{
\begin{figure*}[t]
\epsscale{1.2}
\plotone{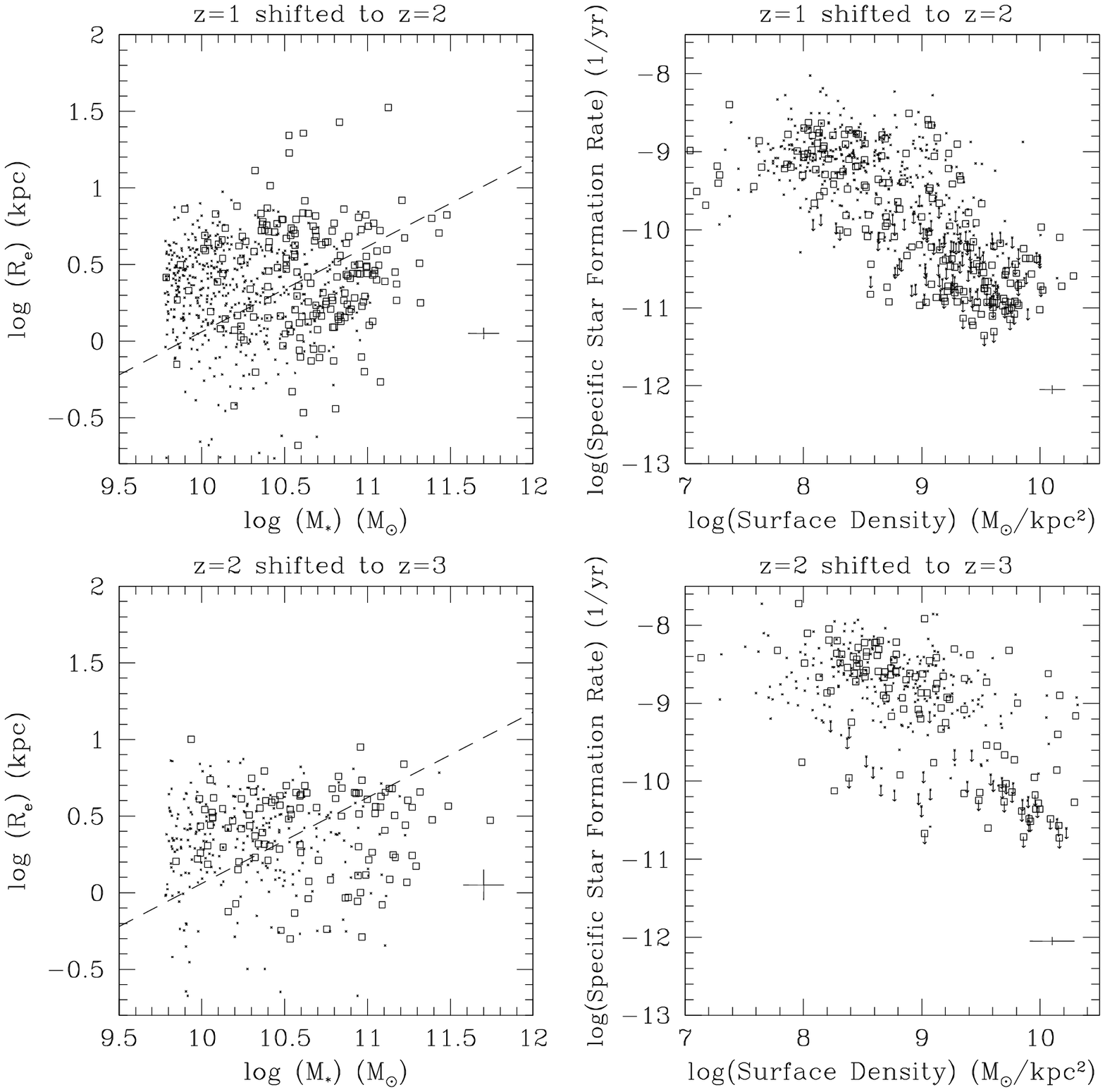}
\epsscale{1.0}
\caption{\small
Simulation  of the selection effects. Top row: galaxies at $0.5 < z <
1.5$ have been shifted to higher redshift by $\delta z =  1$, while
keeping
their intrinsic properties the same.
The open symbols show the galaxies which would still be selected using
the selection criteria in this paper. It is clear that they
are not biased in the mass radius relation or in the relation between
specific star formation rate and surface density.
Bottom row: the same, but now shifting galaxies at $1.5 < z < 2.5$ by
$\delta z = 1$. Again, the resulting relations are the same. If
anything,
the selected galaxies are slightly larger than the original full
distribution,
caused by the fact that small galaxies are red, and therefor dropped
earlier
from the sample. The median radius increases by 15\%.
The simulations shows that the evolutionary effects observed in this
paper are not caused by selection effects.
\label{figone}}
\end{figure*}
}

The dramatic progress in observational capabilities of the last decade
have enabled studies of galaxy evolution and formation to epochs which were
completely inaccessible only 15 years ago.
Nevertheless, despite this dramatic progress, many open questions
remain. Some of these are due to very fundamental problems related to
studies of galaxy formation.
The most important problem is that it has become clear that 
galaxy evolution is a complex process. 
Galaxy mergers change the mass function continuously. 
Furthermore, star formation does probably not occur with nice, smooth
exponentially declining star formation rates -
starbursts  can change the luminosities of
galaxies on a short timescale.
As a result, studies of galaxy evolution and formation are of a
statistical nature: we cannot directly establish which galaxies at $z=1$ would
evolve into what galaxies at $z=0$, and we therefore have to analyze
statistically the sample as a whole.

Modern analyses therefore employ large samples, and attempt to use
stellar masses whereever possible, as these are less likely to evolve
rapidly (e.g., Brinchman \& Ellis 2001, Kauffmann et al. 2003b,
Dickinson et al. 2003, Rudnick et al. 2003, Fontana et al. 2005, 
Borch et al. 2006, Faber et al. 2007).
However, as stressed by Faber et al. (2007), the mass function does
not show a strong mode or other feature. As a result, we have
to characterize the mass function by the 
characteristic mass at which it turns over. The determination of this
characteristic mass and the determination of the general shape of 
the mass function is hard.

A different type of information may come from 
studies of the structure of galaxies. By studying the sizes, colors,
velocity dispersions, etc, one may be able to derive tight
correlations between galaxy properties. These tight correlations can
then be used to study galaxy evolution. For example, by measuring
velocity dispersions, sizes, and luminosities of early-type galaxies
one can establish correlation between mass-to-light ratio and mass and
size of a galaxy (e.g., Djorgovski et al. 1988, Faber et al. 1987).
The evolution of these correlations with redshift (e.g., Franx 1993,
van Dokkum and Franx 1996, van der Wel et al. 2005, Treu et al. 2005,
van Dokkum \& van der Marel 2007) provide the evolution of the 
mass-to-light ratio with redshift, which is essential for a proper
interpretation of the evolution of the luminosity function.
In addition this evolution puts strong constraints on the
evolution of the early-type galaxies per se, and especially the time
at which they formed their stars.
However, we have to add that this interpretation is not entirely
insensitive to the statistical evolution of the class of early-types
(e.g., van Dokkum and Franx 2001). Nevertheless, detailed studies
of galaxy properties can provide important additional constraints on
galaxy evolution, even when relatively small samples are studied
compared
to the full statistical analyses.

In this paper we follow the second approach, where we study a fairly
limited sample of $K$-band selected galaxies in the CDFS 
(1155 galaxies  from Wuyts et al. 2008), augmented
by a local comparison sample based on the Sloan Digital Sky Survey
(Strauss et al. 2002). 
We use size measurements to explore the
relations between color, masses, star formation rates and size.
We show that a good correlation exists between the sizes, masses,
colors and  specific star formation rates of the galaxies from $z=0$
to $z=3$, and we analyze this correlation in this paper.
The paper is built up in the following way: section 2 presents the
data used in the analysis, both the high redshift sample, and the low
redshift sample. Section 3 presents the correlation between mass,
size, color  at redshifts $z=0$, 1, 2 and 3. In addition, the relation
between mass, size and specific star formation rate is presented.
Section 4 presents the evolution of the specific star formation rate as
a  function of redshift, and section 5 presents the evolution of
size. The results are discussed in section 6, and in section 7 we
discuss the potential biases that may occur in current samples of
high redshift galaxies and analyses.
The results are summarized in section 8.

\section{Observations and derived quantities}

\subsection{High redshift ($0.2 < z < 3.5$)}

\figprezero
\figpreone

The high redshift sample of galaxies is taken from the
GOODS-CDF-South field as presented by the
FIREWORKS 
study
by Wuyts et al. (2008). Briefly, Wuyts et al. (2008)
combined 
the available
optical-near IR-mid IR observations on the GOODS-CDF-South field.
The optical imaging consisted of ACS GOODS imaging (Giavalisco et al
2004), and deep imaging with the ESO/MPG 2.2-m telescope 
(Arnouts et al. 2001, Wolf et al. 2004).
The Near-IR imaging was taken with the VLT and ISAAC and
presented in a reduced form by Vandame et al. (2001) and Vandame (in
preparation).
The GOODS IRAC
and MIPS 24 $\mu$m imaging was taken from Dickinson et al. (in preparation).
Wuyts et al. (2008) homogonized the PSFs of the GOODS ACS
and VLT Near-IR imaging data, to derive optical-near-ir photometry,
based on a  K-selected sample.
As the PSFs of the other data sets were significantly worse, the PSF
convolution and fitting technique of 
Labb\' e et al. (2006) and Labb\' e, (in preparation)
was used to derive
colors in the remaining bands. 
The overall photometry derived by Wuyts et al. (2008) agrees well with
that of Grazian et al. (2006), with the exception of an offset
in the IRAC photometry in Grazian et al. (2006), which was removed
in a later update of the Grazian et al. (2006) catalogue (Grazian,
private communication). 
The total effective area of the survey  is 138 arcmin$^2$.

Wuyts et al. (2008) derived photometric
redshifts using EAZY (Brammer et al. 2008), and found good
correspondence between the photometric redshifts and the available
spectroscopic redshifts. The overall difference 
$\delta=(z_{phot}-z_{spec})/(1+z_{spec})$ amounted to 0.03 for the
full sample with spectroscopic redshifts, and was slightly larger for
galaxies with $z_{spec} > 1$: $\delta = 0.05$.

The Wuyts. et al. (2008) sample, and the sample used here,
is selected in the $K$-band.
To assure high quality photometry in the near-IR, only galaxies with a
total $K$-magnitude  below 22.5 were used, and with a signal-to-noise
higher than 10 in the $K$-band. Galaxies with redshifts between
0.5 and 3.5 were used, assuring that the detection band is always
redward of the restframe Balmer/4000 \AA\ break.
The resulting redshift distribution of the sample used in the paper
is shown in Fig. 1.
\figsubzero

Another ingredient in the analysis presented here is the sizes of the
galaxies.
The sizes have been determined in the band redwards of the redshifted
4000 \AA\ break and closest to the rest-frame $g$ band. 
Sersic models convolved with the PSF were fit to each galaxy. 
The sersic index $n$ was allowed to vary between 1 and 4.
The procedure is
identical to that used by Trujillo et al. (2006a), and  Toft et al. (2007).
The procedure was as
follows: for each near-IR tile in the field, the PSF was determined by
averaging the normalized PSFs from the stars in the tile. 
For the ACS imaging, the PSF was similarly determined from stars.
Then
the GALFIT program was used (Peng et al. 2002) to derive the best
fitting position, flux, circularized half light radius $\re=\sqrt{ab}$, 
sersic index, ellipticity
and position angle. Independently, the r4fit program written by the
first author was used to verify the results. This program has been 
extensively used in the past (e.g., van Dokkum \& Franx 1996, van Dokkum et
al. 1998). The results agreed well, with a median systematic offset
as small as 3\%.
We verified that all the results presented later do not change when
the method is changed.  Furthermore, the results change rather little when
the sizes from just the $K$-band imaging are used.
Trujillo et al. (2006a) studied in great detail the possible systematic
effects that can arise from low signal-to-noise, small input sizes,
and other effects. 
In general, the systematic effects are small at high signal-to-noise.
At sizes above 0.1 arcsec, the systematic effects in $\re$ are very
small - which is surprising, as the FWHM of the PSF is about 0.45 arcsec.
As the instrumental setup of Trujillo et al. (2006a) 
was identical to the one used here,
we expect similar errors, and hence we expect that the sizes are reliable
for measured sizes above  $\approx$ 1 kpc (corresponding 
to 0.12 arcsec at $z=2$).

Galaxy masses have been estimated from SED fits to the full
photometric dataset shortward of 24$\mu$m, 
and are presented in F\"orster Schreiber et
al. (in preparation)
following similar procedure as described by F\"orster Schreiber et al.
(2004).
The masses used here are based on fits with stellar
population models by Bruzual \& Charlot (2003). The models used in the
study were based on star formation histories with exponentially
declining
star formation rates. The time scales were 0 (single stellar
population),
300 Myr, and infinity (constant star formation).
A Calzetti et al. (2000) extinction curve was used, and the extinction in the
V band $A_V$ was allowed to vary between 0 and 4.
The single stellar population models did not include dust extinction.
The fits also produced
estimates of the star formation rates, and these are briefly used
below for comparison.
The IMF used was the Salpeter (1955) IMF. The results were normalized
to a Kroupa (2001) IMF, by multiplying the masses and star formation
rates by $10^{-0.2}$. We verified that this approximation is accurate
to a few percent.
We notice that the masses presented here decrease on average by 1.4
when Maraston (2005) models are used, without a dependence on
redshift (see also Wuyts et al. 2007b).
\figpretwo

Star formation rates were also estimated from the UV and 24 $\mu$m MIPS fluxes.
First, mid-IR fluxes were estimated using the models by
Dale \& Helou (2001), and the prescription by Labb\' e et al. (in preparation)
and Wuyts et al. (2008). In short, a large range of models was used
to convert the MIPS flux to a bolometric flux, and the mean of the log
of the bolometric fluxes was used as a best estimate. Typically, the
top and bottom estimates varied by a factor of 3-5 from our best
estimate.
We note, however, that Wuyts et al. (in preparation) found that the 
conversion used
here agreed within 10\% with the conversion given by Papovich et
al. (2007),
based on observed 70 $\mu$m and 160 $\mu$m fluxes for galaxies in the
E-CFDS.
The star formation rate of the galaxies was estimated assuming the
Kroupa IMF : $SFR = 0.98\ 10^{-10} (L_{IR} + 3.3 L_{2800})$.
This is the relation by Bell et al. (2005), and it is based on the relations
given by Kennicutt (1998). The Bell et al. (2005) 
relation has been adapted to the Kroupa IMF used in this study.

The star formation rates have been measured for all galaxies,
and the 1-sigma errors on the measured star formation rates dependend 
on redshift. The typical random errors are 
less than 1 $\Msun$ year$^{-1}$ at $z=1$, 
less than 5 $\Msun$ year$^{-1}$ at $z=2$, 
increasing rapidly to 25 $\Msun$ year$^{-1}$ at $z=3$.

\figsubone
We compare the star formation rates derived from the UV+ 24 $\mu$m MIPS
fluxes
with the star formation rates derived from the SED fits in
Fig. 2. We compare the star formation rates at different redshift
intervals, $z\approx 1$, $z\approx 2$, $z\approx 3$.
We can see a good correlation in the first two redshift bins.
There is a small offset at $z\approx 1$, with a median 
of SFR(UV+MIPS)/SFR(SED) = 0.77.
At $z\approx 2$, the median ratio is 1.29.
These deviations are much smaller than the uncertainties in the 24
$\mu$m MIPS flux to total IR flux conversion, which are up to a factor
of 3.
At $z\approx 3$, the correspondence is less good. This is
expected:
observational errors on the 24$\mu$m fluxes become more important,
and the 8$\mu$m  PAH emission feature 
shifts out of the 24$\mu$m band at those redshifts.
Overall, the agreement between the star formation rates from UV+MIPS and
SED-fits is surprisingly good at $z\le 2.5$.
This may come as a
surprise, as other authors found that the MIPS based star formation
rates are significantly too high for a substantial fraction of the galaxies
(e.g., Daddi et al. 2007). We do not find
such an effect. This may be due to the fact that we use a linear 
conversion from MIPS flux to
full IR flux. This apparently works well.
The assumption is further
justified by the work of Papovich et al. (2007), who found that the
earlier empirical conversions of MIPS flux to star formation rate for $z=2$
galaxies generally over-predict the star formation rate. As a matter
of
fact, the results of Papovich et al. (2007) are in very good agreement
with a simple linear conversion.
However, we have to note that our estimates of the star formation
rates
may still be wrong: we have
not measured the full IR flux and depend on extrapolations.
We remark, however, that Daddi et al. (2007) concluded that the
star formation measured from the dustcorrected UV is a good representation
of the true star formation rate. Our SED fits should be a good
approximation of those values. 
But at the same time, we find that our UV+MIPS star formation rates
correspond well to the star formation rates based on SED fits.
Hence, in the rest of the paper, we use the UV+MIPS
star formation rates.
Galaxies with X-ray detections (Giacconi et al. 2002) were omitted
from the sample, as these have likely AGN.

For consistency, we use the MIPS derived star formation rates also at $z>2.5$.
We note that at $z\approx 3$, the MIPS star formation rates are higher than
the SED derived star formation rates by a factor than 2.8 (Fig. 2).
We verified that the results obtained in the paper remain valid when using the
SED derived star formation rates. The main difference is a systematic 
decrease in derived star formation rates and specific star formation rates
at $z=2.5$ and higher.

\subsection{Low redshift sample: SDSS}

For comparison at low redshift, we use the SDSS sample, as used by
Kauffmann et al. (2003a,b). These authors derived masses for galaxies
from the DR2 release 
(Abazajian et al. 2004). The masses are based on 
mass-to-light ratios estimated from analysis of the nuclear spectra.
We applied a small correction to the mass-to-light ratios: 
as the colors of the nuclei
are generally redder than the color of the galaxy as a whole, the
mass-to-light ratio is generally slightly overestimated. We corrected
the mass-to-light ratios with a factor of $10^{1.7 \Delta(g-r)}$,
where $\Delta(g-r) = (g-r)(petrosian) - (g-r)(fiber)$.
Furthermore, we corrected the masses by the fraction of light missed
in the Petrosian aperture. We used the sersic fits by Blanton et
al. (2003) to derive the total fluxes, and the ratio of the total
flux to the Petrosian flux.

The rest-frame colors of the galaxies were derived using INTEREST
(Taylor et al. 2008, in preparation), 
and are determined in a way consistent with the
high redshift sample.
INTEREST determines the rest-frame colors and fluxes from the
observed magnitudes, using the algorithm defined by Rudnick et al. (2003).
The star formation rates were taken from Brinchmann et al. (2004).
Finally, the sizes (half-light radii) were taken from Blanton et
al. (2003).

In the rest of the paper, we use the sample between redshifts of 0.05
and 0.07. The lower limit is used to avoid a strong bias against
massive galaxies, resulting from the magnitude limit imposed on the
sample with masses from Kauffmann et al. (2003a).
The high redshift limit is used to avoid the worst selection effects
against apparently small galaxies, and to avoid large uncertainties 
in the derived sizes.
This sample contains 21722 galaxies.
The sample includes both star-forming galaxies and
galaxies without star formation.

\section{ Correlations between size, mass, color and star formation rates}

\subsection{Mass-color and Mass-size relations }

\figsubtwo
We can now start to analyze the correlations between galaxy parameters
out to $z=3.5$.
We start by showing the well known correlation between color and
stellar mass (e.g., Kauffmann et al. 2003a, Borch et al. 2006).
The top row in Fig. 3 shows the result  for the galaxies
in our sample. Here, and in following plots, we divide the galaxies in
4 bins: low redshift ($0.05 < z < 0.07$  from SDSS), $0.5 < z < 1.5$, 
$1.5 < z < 2.5$, and
$2.5 < z < 3.5$, the latter all from the CDFS.
Fig. 3 shows that  there is a correlation between color and mass
out to $z=3.5$:
low mass galaxies being generally blue,
and high mass galaxies being red. At low redshift we see the well known
red sequence extending to high masses. 
At high redshift we lack the accuracy to establish the red sequence,
but a correlation between color and mass probably persists as massive 
blue galaxies are scarce.
These results agree well with
those of Kauffmann et al. (2003a), 
Borch et al.  (2006) for galaxies at $z < 1$.
\figprethree

\figprefour

As the redshifts used for this figure are mostly photometric
redshifts, the errors in the colors are significant at higher
redshift.
Hence the absence of a tight red sequence at $z=2$ can be caused
entirely by the observational errors.
We note that Kriek et al. (2008) found a  red sequence at $z=2.3$ for
a sample with spectroscopy.
It is striking that at $z=2$ and above, there appears to be a lack of
low mass, red galaxies. 
We have to realize, however that red, low mass galaxies are 
progressively missed from the
samples, as they have low mass-to-light ratios, and drop out of the $K$-band
selected samples first. We have indicated in Fig. 3 the  lines of
75\% completeness.
We determined this in the following way. 
We divided the galaxies in a redshift bin into separate color
bins. In each color bin, the galaxy fluxes and masses were scaled downward so
that the resulting signal-to-noise in the K band was at our selection
limit.
For each galaxy, the resulting mass is the limiting mass at which that 
particular galaxy could have been observed. 
Finally, we determined the mass limit at which 75 \%
of the galaxies were detectable, and show it in Fig. 3 with the
dashed curve.
It is clear that the absence of red, low mass galaxies at $z=2$ and
$z=3$ is mostly due to the incompleteness of the sample.

Whereas a general relation exists between mass and color, 
it is noticeable that at stellar masses between  $10^{10}$ and 
$10^{11}$ $\Msun$,
the spread in color is very large, up to 0.8 magnitude in $u-g$.
This is not only the case at low redshift, but also at higher redshifts.
As the color is a function of stellar age, it implies a large range
in star formation histories of galaxies at a given mass.
Other indicators (D4000, specific star formation rate
\footnote{The specific star formation rate is the star formation rate
  divided by the stellar mass.}
derived from emission
lines) show the same, large spread at these intermediate masses in the
local universe (e.g., Kauffman et al. 2003b, Brinchman et al. 2004).
Clearly, mass is not the only factor determining the star formation
history,
and galaxies are not uniform in this mass interval.

Next we show the relation between mass and effective radius in the bottom
row
of Fig. 3.
There is a well defined trend at low redshift, with high mass 
galaxies being larger than low mass galaxies.
The trend is weaker at high redshift, but this might be caused by
selection effects, as small galaxies are preferentially missed first
(see Appendix A).
Just as for the mass-color relation, we find a significant scatter
at a given mass at all redshifts.
This was found earlier by Shen et al. (2003) for the SDSS sample,
and it persists to high redshift.
Shen et al. (2003) found that the rms in log radius is 
0.3-0.5 dex at a
given mass, and the distribution is log-normal at 
low redshift.

\subsection{ Mass-size-color relation}

If we take these two results together, we conclude that 
galaxies at masses between $10^{10}$ and
$10^{11}$ $\Msun$
have a large spread in
effective radius, and a large range in colors, from $z=0$ to $z=3$ !
This raises the simple question whether a different, underlying parameter
might cause this variation in color, and size.
This might be expected if, for example, the galaxies
consist of large, blue disks, and small, red bulges, and if the
bulge-to-disk
ratio varies, at a given mass.

\figsubthree

To investigate whether an additional parameter may cause the scatter, 
we present the relation between
color and effective radius in narrow mass bins.
Figure 4 shows these  relations
in our 4 redshift bins.
Interestingly,  we find a  tight relation between color and
effective radius, showing that the {\em combination} of mass and size can
predict the color very well.
The relation persists to $z=2$, with too few galaxies at $z=3$ to
establish it at that redshift.
The average scatter in $u-g$ color around a simple fit to color-size relation
at a given mass
is very  small at 0.16 mag 
in the mass bins between $10^{10}$ and $10^{11}$ $\Msun$.
This is a significant reduction of the scatter, which is 0.33 mag in
$u-g$ color in the same mass range if we do not correct for the
color-size relation.
The scatter listed here, and in the remainder of the paper is measured
with the normalized Median Absolute Deviation (nMAD). This is the
median absolute deviation multiplied by 1.48 so that the nMAD of
a gaussian is equal to its dispersion.

This result suggests that indeed a single underlying parameter might
drive the variations between galaxies.
Hence we test  whether the relation
between color, mass, and size can be written in a simple form.
We assume a relation of the form

$$ u-g = a ( \log\ Mass - b \log\ r_e ) + c,$$

and we derive the values of $a$, $b$, and $c$ which minimize the scatter.
The selection effects discussed earlier make it hard to do this
test unambiguously at high redshift.
At low redshift ($z=0$ and $z=1$), we find that the scatter is
minimized for $b$=1.05 and $b=1.55,$ respectively, with a value of $b=2$
giving a similar, but slightly higher scatter. At $z=2$, selection
effects are already quite important, but we still find a similar value
of $b=0.89$. At $z=3$, we find that $b=0.28$ produces the lowest scatter,
likely caused by the selection effects against low mass, red galaxies,
making it impossible to establish the mass-color-size relation at that
redshift.

We note that
the coefficient $a$ has no special physical meaning, as it is dependent on
the color used on the left side of the equation.

However, the coefficient $b$ has a special meaning, as it can indicate the
physical parameter underlying the relation.
Our main result is that $b$ is close to 1-1.5 for most redshifts
($z=0$ to $z=2.5$). This suggests that the relevant parameter is 
$Mass/\re$, related to velocity dispersion,  or
$Mass/\re^2$, related stellar surface density.
In the following, we use the term ``inferred velocity dispersion''
for the quantity derived from $M/\re$,
to make clear that it is not the true stellar velocity
dispersion for the galaxies. The two agree well at low
redshift, see, e.g., Drory, Bender, \& Hopp (2004), but at high
redshift this remains to be demonstrated, especially for gas rich
galaxies.

\figprefive

This correlations obtained here are similar to the results 
of Kauffmann et al. (2003b,
2006)
for the SDSS
sample, who noted
that the specific star formation rate correlates better with stellar surface
density than with mass.
The scatter in $u-g$ color is low, for both the color-inferred velocity
dispersion and color-surface density relations: 0.21 and 0.22 mag
respectively, 
for our bins at $z<2.5$.

We emphasize that the single correlation with surface density or
inferred velocity dispersion
 is unlikely to be complete: as can be
seen in Fig. 4, at $z=0$, at each mass interval, a ``narrow red sequence'' 
exists
over a  range in radii. The relation above does not represent this
sequence properly, and the relation has less predictive power than, for
example, the Fundamental Plane relation for early-type galaxies alone.

Given this result, we now show the relation between color and surface
density in the left panels of Fig. 5.
It is striking that a well defined relation exists at all redshifts,
with red galaxies having high surface density, and blue galaxies
having low surface density.
The relation exists at all redshifts, but we have to note that
at $z=3$ the number of points is low, and the relation is weak.
More, and deeper data at high resolution
would be needed to establish the relation better
at that redshift.
\figsubfour

The relation between color and inferred velocity dispersion is shown in the
right panels of Fig. 5.
We calculated the inferred velocity dispersion from $\sigma = \sqrt (0.3 G
M/\re)$,
where the constant has been chosen so that the inferred velocity
dispersions of the SDSS galaxies match the measured dispersions well.
Again,  we  find a well defined relation between the inferred velocity
dispersion and the color,
with somewhat smaller scatter  in color than for the relation between
surface density and color.
 
\figpresix

\subsection{Specific star formation rates as a function of surface
density}

The color variations found above are interesting by themselves, but 
the interpretation is complex, as color is a function of both
star formation history and dust.
A more direct diagnostic of the star formation history of the galaxies
is the specific star formation rate. 
The specific star formation rate is the star formation rate divided by
the mass, and it is the inverse of the time it would take to form the
galaxy
if the star formation rate were constant.
In general, the specific star
formation
rate and color of galaxies are well correlated, and the correlation
between surface density and color suggests therefore that a similar
correlation may exist between surface density and specific star
formation
rate.

\figsubfive
Figure  6 presents the specific star formation rates plotted 
against the surface density and inferred velocity dispersion of the galaxies.
As can be seen, good correlations exist.
The  high surface density galaxies have generally lower specific 
star formation rates, and the low surface density galaxies have
generally high specific star formation rates.
We note, however, that the scatter is larger for the SDSS sample
than when the color
is plotted against surface density, and there is no clear linear 
relation.
Either aperture correction effects play a role (Brinchmann et al. 
2004), or a correlation
between dust and specific star formation rates produces
lower scatter in  color than in specific star formation rate
by itself.
At higher redshift, this difference is not strong.

We conclude that the relation between color and surface density is
directly related to the relation between specific star formation rate
and surface density. 
This result extends the earlier analysis of Kauffmann et al. (2003b, 2006)
to significantly higher redshift.

\section{Evolution of specific star formation rate with redshift}

The tight correlation between color and specific star formation rate
with surface density and inferred velocity dispersion suggests that
evolutionary studies should focus on using these parameters for
studying the evolution with redshift - in contrast to using
mass (e.g.,  Cowie et al. 1996, Brichmann \& Ellis, 2000, Juneau et al. 2005)
Below, we  explore the evolution of specific star formation
rate as a function of mass and of  surface density.

\subsection{Specific star formation rates in bins of mass}

It is now well established that galaxies generally had higher star 
formation rates at higher redshift (e.g. Lilly et al. 1996,
Cowie et al., 1996, Bell et al. 2005).
The evolution of the star formation rate is generally found to be
dependent on mass: the most massive galaxies stopped forming stars
the earliest (also known as down-sizing, e.g., Cowie et al. 1996, 
Brichmann \& Ellis, 2000, Juneau et al. 2005, Zheng et al. 2007)

\figsubsix

Fig. 7 shows the evolution of the specific star formation rate in
narrow
mass bins.
The specific star formation rate is the star formation rate divided
by the stellar mass. It is the inverse of the time in which the galaxy 
would have formed, given the observed mass and star formation rate.
We can see in Fig. 7 that the specific star formation  rates increase
with increasing redshift, consistent with previous results (e.g.,
Juneau et al. 2005, Zheng et al. 2007, and references therein)
The increase can be seen in all mass bins. 

The specific star formation rates increase by a large factor:
whereas at $z=0$ the highest
specific star formation rates range around $10^{-10}$ \yrinv,
by $z=1$ the maximum specific star formation rates increase to
$10^{-9}$ \yrinv,
and by $z=2$ to even higher values ($10^{-8.5}$ \yrinv). 
In this discussion, we have to keep
in mind that the specific star formation rates suffer from systematic
uncertainties.
Nevertheless, it is striking how strongly the specific star formation rates
increase with redshift, and we notice that the trends at $0.2 < z < 1.5$
extrapolate at $z=0$ to values consistent with the local SDSS values.

\figpreseven

We also find a trend with mass: in the lowest mass bins, the majority
of
the galaxies have already very high specific star formation rates by $z=1$,
whereas at the highest masses, the galaxies with high star formation
rates start to dominate at significantly higher redshift.
These results are very similar to those obtained before by many other
authors (e.g., Cowie et al. 1996, Brinchmann \& Ellis 2000, 
Zheng et al. 2007 and references therein), and are generally
characterized as ``down-sizing''.

However, we notice that all mass bins contain galaxies with very low
specific star formation rates to very high redshifts.
We have to keep in mind, that the upper limits on the specific star
formation
rates at $z=2$ are quite high, 
especially
for the lower mass galaxies. 
This is simply caused by the limited depth of the MIPS 24 $\mu$m imaging:
Even though the MIPS exposure is extremely deep,
the limiting depth produces rather high limits on the specific star
formation
rate for low mass galaxies.

Additionally, we note that the highest specific star formation rates occur for
the lowest mass galaxies. These do not have the highest star formation
rates in an absolute sense:
a specific star formation rate of  $10^{-8}$ \yrinv\ at $\Mstar = 2\ 10^{10} \Msun$
corresponds to a star formation rate of 200 $\Mstar$ \yrinv, whereas
a specific star formation rate of $3\ 10^{-9}$\yrinv\ at $\Mstar = 8\ 10^{10} \Msun$
corresponds to a star formation rate of 240 $\Mstar$ \yrinv.
Obviously, characterizing  galaxies by their absolute star formation rate
over-emphasizes massive galaxies, which are actually not as
extreme as the lower mass galaxies.

\figsubseven
\subsection{Specific star formation rate in bins of surface density}

The wide range in specific star formation rate in the mass bins is a
striking
feature in Fig. 7.
In Fig. 8, we show the specific star formation rate in bins of surface density.
It is striking that at low surface density,  the specific star formation
rates are always high, and at high surface density, the specific 
star formation rates are generally low; but at intermediate surface densities,
($10^{9}$ - $10^{10}$ \msunkpc)
we see a transition from low specific star formation rates at low redshift.
to high specific star formation rates at high redshift.
At a surface density of $10^{9} -  10^{9.5}$ \msunkpc\ this transition takes
place
in a fairly narrow interval between $z=0.6$ and $z=1.6$, and at
the
surface density interval of $10^{9.5}-10^{10}$ \msunkpc, it takes
place 
around $z=1.5-2.5$.

By comparing Figs. 7 and 8, we conclude that the evolution is 
better  defined by the surface density of the galaxies, than by their
mass. At all masses and nearly all redshift, we find galaxies with
very high specific star formation rates, and galaxies with very low
specific star formation rates. When galaxies are sorted by surface
density,
we see that low surface density galaxies have high specific star formation
rates
at all redshifts, high surface density galaxies have low specific 
star formation rates at nearly all redshifts, and we find a clear transition
zone for intermediate surface density galaxies.

\subsection{The evolution of threshold surface density with redshift}

Kauffmann et al. (2006) analyzed a sample of local SDSS galaxies, and 
introduced  a threshold surface density, below which galaxies have
nearly constant specific star formation rates, and above which the
specific star formation rate declines rapidly.

Our results imply that similar threshold surface densities
can be defined at higher redshifts, and that the evolution of the
galaxies
can be well described by an increase in the threshold surface density
with increasing redshift.

\figpreeight
\figprenine

\figsubeight

We define the threshold surface densities in the following way:
we determine the median specific star formation rates in surface
density bins of width 0.5 dex,
sampled at 0.1 dex. 
The results are shown for $z=0,1,2$ in Fig. 9.
We find that the median specific star formation rates are fairly
constant at low surface densities, and decline rapidly at higher
surface 
densities. 
The threshold surface density is defined by requiring that the specific
star formation rate is 3 times lower than the median at the low
surface density end.

We determine the thresholds in redshift bins of width $\Delta z=1$ for
our CDFS sample, and we sample them at redshift steps of 0.5.
The results are shown in Fig. 10. The left panel shows the threshold 
surface density versus  redshift.
As can be seen, the threshold evolves quite strongly with redshift - 
proportional to $(1+z)^{1.5 \pm 0.12}$. 

\figsubnine

This evolution by itself would obviously cause a strongly increasing star
formation rate density with redshift, as more and more galaxies fall below
the threshold. However, it is not the full story, as we can also see that
the specific star formation rate below the surface density threshold 
increases rapidly with redshift
(Fig. 10b). The evolution is fast at low redshift, proportional
to $(1+z)^{3.8 \pm 0.2}$.

\section{Evolution of mass-size relation with redshift }

We saw above that the specific star formation rate is strongly
correlated with the surface density of the galaxies, and evolves
strongly at a given surface density.
This is not the only evolution taking place: the surface
densities and sizes of galaxies at a fixed mass are also expected to evolve.
This evolution has been studied before (e.g., Trujillo et al. 2006a, and
references
therein). We revisit this issue here as our sample size is significantly
larger at high redshifts.
To be consistent with existing literature, we study the mass-size
relation.
The relation between surface density and mass follows directly from
using $\Sigma \propto M/\re^2$.

We note that the evolution of the mass-size relation
is  a valuable diagnostic of the evolution of
galaxies:
in simple models of disc formation, the sizes of discs are assumed to
evolve at the same rate as the halo size (e.g., Mo, Mao, \& White 1998).
This has
been used to predict the evolution of the mass-size relation for
such galaxies (e.g., Mo et al. 1998). More complex models find
generally weaker evolution (e.g., Somerville et al. 2008).
The evolution of the mass-size relation for spheroidal galaxies is
expected to evolve even faster 
(e.g., Kochfar \& Silk 2006, Hopkins et al. 2007,2008)

\figpreten
\figsubten

The first studies of the evolution of the size-mass relation to $z=3$
were done by Trujillo et al. (2006), based on ISAAC imaging data
on the smaller but deeper fields of the FIRES survey (Franx et al.
2003, Labbe et al. 2003a, Forster Schreiber et al. 2006).
The sample studied here is based on imaging with the same instrument, but
on the much larger CDF-South field.
Fig.  11 shows the evolution of the radii of galaxies with redshift in
narrow mass bins. As we can see, significant evolution is present.
On average, the sample produces an evolution of $\delta \log \re =
-(0.13\pm 0.02) z$,
or $\re \propto (1+z)^{-0.59\pm 0.10}$. This is an average of the
evolution seen in the mass bins with $M_* > 2.5\ 10^{10}$, where the
sample is more than 70 \% complete at the $z=2$ bin.
The evolution may still be an
underestimate, as the lowest mass bins are still deficient in small, red
galaxies at high redshift ($z \ge 2.5$).

As can be seen in Fig. 11, the evolution is fastest for galaxies with
masses above 6.3 $10^{10}$ $\Msun$. These give $\re \propto
(1+z)^{-0.71 \pm 0.07}$.
Again this difference may be partly or fully due to incompleteness at the lower
masses:  the smallest galaxies are typically red,
and they drop out of the $K$ selected samples earlier than the larger, 
blue galaxies.
Deeper data will be required to verify this trend.
Furthermore, we note that the smallest galaxies are small with regards
to the PSF, and hence the size measurements around 1kpc and below should be
considered to be uncertain (see also section 2).
Hence higher resolution imaging is needed to verify those sizes.
We  note that the $z=0$ SDSS size measurements agree well with the
trend found for the full dataset. There is no indication that the
SDSS measurements are biased compared to the low redshift CDFS measurements.

The evolution found here agrees well with that found by Trujillo et al. (2006a),
who found an average evolution for the full sample 
of a factor of 0.48$\pm$ 0.05 out to z=2.5, compared to the evolution of
a factor of 0.48$\pm$0.08 found here (derived from our average 
redshift evolution).

Given the strong evolution, one may wonder to what extent these
results could be affected by serious systematic errors in establishing the
stellar masses of galaxies. We note that the evolution is so strong, that
it cannot be caused by an error of a factor of 3 in the stellar mass at $z=2$.
The typical size of galaxies with mass of 6.3-10 $10^{10}$ $\Msun$
in our sample is 1.6 kpc at $z=2$. If we had overestimated the mass
by a factor of 3, we should compare their size with galaxies at z=0 of a
mass of 2.1-3.3 $10^{10}$ $\Msun$. 
As we can see in Fig. 11, galaxies with that mass
have a size of 2.5 kpc at $z=0$, implying an evolution with a factor
of 0.64.
Hence it seems rather unlikely that systematic
errors in the mass determination cause the observed evolution.
Incompleteness could be another factor, as we may miss the larger
galaxies
more easily
by surface brightness effects. This is straightforward to test by simulating
the effects of such selection on the observed low redshift galaxies, by putting
them at higher and higher redshift, and simulation the selection effects.
In appendix A we show that this effect
is almost negligible, which is due to the fact that the PSF is significant
compared to the sizes of the galaxies. Hence we miss rather few large galaxies,
and we cannot easily explain the absence of massive, small red galaxies at 
low redshift.
The most significant bias is to miss low mass, red galaxies at high redshift, 
because they are fainter in $K$, 
and this biases the sample towards larger galaxies at the same mass.
Hence the true evolution of the mass-size relation may even be faster
than the evolution observed here.

The scatter in the mass-size relation at $z>0.2$ is approximately 0.3 dex
in log radius. This is comparable to the scatter in the relation for
the SDSS sample for galaxies with masses lower than $ 4\ 10^{10}$
$\Msun$ (Shen et al. 2003). The scatter is comparable to the evolution
found here, showing that it is possible that some individual galaxies
have no evolution in either parameter. The largest galaxies at $z=2$,
for example, are consistent with galaxies at $z=0$. The small, massive
 galaxies at $z=2$ are not consistent with $z=0$ galaxies, a point
made also by many other authors (e.g., Cimatti et al. 2008, van Dokkum
et al. 2008, and references therein).
Therefore, the population as a whole must evolve.

\section{Implications for galaxy evolution}

\subsection{General evolution and transformation onto the red sequence}

We have found that at all redshifts, color and specific star formation
rates correlate well with surface density and inferred velocity
dispersion (which is $\sqrt{0.3 G M/\re}$).
The lower the surface density and inferred velocity dispersion, 
the bluer the colors,
and the higher the specific star formation rates.
The implication of this result is simply that the high surface
density galaxies are older, have low star formation, and must have
formed their stars earlier than the low surface density galaxies.
This had been found earlier by Kauffman et al. (2003b, 2006) at low redshift,
and these new result show that the relations persist to at least $z=2.5$.

The evolution in the relations is as expected: at higher redshifts,
high specific star formation rates are found at higher surface
densities/
inferred velocity dispersions than at lower redshift.
This is consistent with a simple picture in which  the high surface density 
galaxies which are dead at low redshift were forming stars at some 
higher redshift. The apparently smooth increase  of the characteristic
surface density at which high star formation occurs suggests that the 
star formation
history is a simple function of the surface density of the galaxy.
This is quite striking as the star formation rate of a galaxy is
driven by complex processes like mergers, gas accretion, and other
processes which will vary with time.
Hence these processes affect both the star formation rates and the
surface
densities to result in a fairly simple relation
between surface density and star formation history.

Interestingly, the results provide independent evidence that many galaxies 
which are on the red sequence at $z=0$ were not "red and dead" at $z=1$:
their typical surface densities are around $10^9\ -\ 3\ 10^9 $
\msunkpc\  at $z=0$, and
approximately half of the galaxies with such surface densities at $z=1$ are
forming stars (see Fig. 8).
We note that this evidence is completely independent of the evidence based
on the mass density evolution of the red sequence galaxies, which has shown
that the mass density has increased significantly between $z=1$ and $z=0$
(e.g., Bell et al. 2004, Faber et al. 2007). The evidence presented here shows
that the galaxies with the structure of red-sequence galaxies at $z=0$ are
generally forming stars at $z=1.5$. Of course, as galaxy evolution is a
complex process which may include both merging and star formation, we cannot
uniquely identify what the progenitors are of $z=0$ red galaxies once we
allow for structural evolution.
However, we can confidently exclude the possibility
that all red sequence galaxies at $z=0$ have passively evolved from $z=1$ red
galaxies: many of the galaxies with the corresponding structural 
parameters at $z=1$
are not dead.
The evolution of the galaxies onto the "dead" zone of surface density$> 10^9$ 
\msunkpc\ 
likely involves both star formation and merging: the first from
the direct observational evidence presented here that the galaxies
with surface density $> 10^9$ \msunkpc\ were forming stars at higher
redshift,
the latter from theoretical
predictions, and determinations of the merger rate 
(e.g., Bell et al. 2006, van Dokkum et al. 2005), 
which are still quite uncertain, however.

The steady increase with redshift of the threshold density 
(above which the specific star formation rates drop)
suggests that similar processes were at play between $z=1$ and
$z=2$, but then at higher surface densities/inferred velocity dispersions.

\subsection{Properties of quiescent galaxies out to $z=3$}

One of the striking features is that quiescent galaxies exist at all redshifts.
At all redshifts, they are the galaxies with the highest surface densities,
and the highest inferred velocity dispersions. However, their sizes at $z=2$ and above
 are
much smaller than their sizes at $z=0$, and their surface densities much
higher. This had been noticed before
by many authors (e.g., Trujillo et al. 2003, 2006a,b, Daddi et al. 2005,
Toft et al. 2007, Cimatti et al. 2008, van Dokkum et al. 2008). 
The results here indicate that
their small sizes are related to the general evolution of the
mass-size relation. 
When we select only quiescent galaxies by requiring
that the specific star formation rate is smaller than $0.3/t{hubble}$,
we find a size evolution
of $\re \propto (1+z)^{-1.09 \pm 0.07}$, 
significantly faster than the evolution for the full
galaxy sample. 
This is measured for masses larger than $4\ 10^{10}$ $\Msun$, where
good limits on the specific star formation rates are achieved.
The error is the formal error from comparisons between the mass bins.
In reality, the error is larger as many of the galaxies have very
uncertain
sizes below 1 kpc.
However, we note that van Dokkum et al. (2008) used high resolution NIC2
imaging
and found a similar evolution
for massive quiescent galaxies:  a factor of 5 between $z=0$ and
$z=2.3$.
Hence the evolution of the quiescent galaxies is faster than
the full galaxy sample, by a factor of two or more.
Given our selection effects at low masses, and our limited resolution,
deeper studies with better resolution are needed to measure the
evolution more accurately.

The fact that quiescent galaxies exist out to the highest redshift is 
very significant,
as it shows that the mechanism which shuts off star formation
in galaxies was already present at high redshift ($z=3$ and above).
In many ways, it does not come as a surprise that the quiescent
galaxies are small: once 
star formation shuts off, they apparently do  not continue to accrete 
cold gas in their outer parts which would make them grow in size.
Hence their size remains fixed, whereas the star forming galaxies
grow in size.
However, it is surprising that the $z=2$ quiescent galaxies are much
smaller than any quiescent galaxies in the nearby universe
with reasonable number densities.
Hence, in one way they must have "disappeared", or surveys of the
nearby universe are incomplete.

Incompleteness in the SDSS survey certainly does play a role: due to the
star-galaxy separation criteria used, several known compact galaxies from the
7 Samurai survey (Faber et al, 1989) are missing in the SDSS survey.
These are galaxies like NGC 4342 and NGC 5845, which have velocity
dispersions above 200 km sec$^{-1}$, and sizes below 0.5 kpc. Such
galaxies are extremely rare in the SDSS (less than 1 in 10$^4$).
The volume in which the compact, massive galaxies can be found is
rather small in the SDSS, 
as they are excluded when at small distance because they
are too bright in the fiber aperture, and they are excluded at large distance
because they are too small (Strauss et al. 2002). 
Despite these potential problems, it is clear that evolution also takes
place between $z=2$ and $z=1$, where such incompleteness should play no role.
The simplest explanation is that the small galaxies grow by merging with
larger galaxies, whether star forming or not. As all other galaxies 
are larger, any merger is expected
to increase the size. Furthermore, additional accretion of gas and
star formation in the outer
parts would also scale up these galaxies.

\subsection{ A simple accretion model to explain why blue galaxies are large}

The fact that blue galaxies are larger than red galaxies of the same
mass may find a very simple explanation. In the nearby universe, we know
that star forming galaxies with masses between $10^{10}$$\Msun$ and
$10^{11}$ $\Msun$ are
generally multi-component: a red bulge, with a blue disk around it.
As a matter of fact, the Hubble sequence is correlated with
bulge-to-disk ratio, and is correlated with color, and 
the sequence satisfies the same general
trend as the trend observed here. Hence a simple explanation for the
trend between  specific star formation rate and surface density is
that the galaxies with significant star formation have accreted gas,
which forms stars preferentially in the outer parts, 
and this why they are both larger, and bluer.
This gas accretion is assumed to be fairly regular and 
possibly caused by minor mergers which do not stir up the galaxy.
For some reason, gas accretion and star formation are shut off for
high density galaxies, either due to AGN (e.g., Croton et al. 2006,
Bower et al. 2006), or heating of the gas due to 
shocks occurring naturally in massive, forming galaxies
(Dekel \& Birnboim 2006, Birnboim et al. 2007,
Naab et al. 2007).

The persistence of the relation at higher redshift, but then shifted to
higher surface density, suggests that similar processes occur out to $z=2.5$
and beyond. The fact that the threshold surface density is  higher
may be a simple result from the fact that the halos of galaxies are
smaller at high redshift, and hence the galaxies are denser. 
Galaxy size is correlated with the star formation history, as the
galaxies which have accreted material recently are expected to be larger,
and have higher specific star formation rates than those who have not accreted.
In short, a very simple picture is one in which pre-existing star
forming galaxies accrete  material preferentially in the outer parts,
where stars are formed.
For some reason, high surface density galaxies do not accrete such material.
In this picture, star forming galaxies with $M > 10^{10}$ $\Msun$ at 
$z=2$ are very analogous to 
$z=0$: old centers, younger outer parts. This specific prediction of
gradients can be tested with higher resolution imaging data, as would
be provided, for example with WF3 on HST.
The imaging study by Labb\'e et al (2003b) showed evidence for
substructure and gradients for a sample of large galaxies to $z=2$.
As the angular momentum of the accreted material is thought to set
the scale of the galaxies, it is also natural to expect that the blue 
galaxies may have disks of star forming gas, analogous to the low 
redshift galaxies. There is at least some evidence for this from
kinematic studies of high redshift galaxies
(e.g., F\" oster Schreiber et al. 2006, Genzel et al. 2006, 
Wright et al. 2007, Law et al. 2007)
As we  see below, and as noted in these  kinematical studies,
the nature of the star forming galaxies at $z > 1.5$ 
may be significantly different from what
we call "disk galaxies" in the nearby universe.

\subsection{The enigmatic nature of strongly star forming galaxies at $z \ge 1.5$.}

As we have seen, at $z\ge1$ the strongly star forming galaxies have
very high specific star formation rates, well above $1/t_{hubble}$.
For example, at $z=1$ the typical specific star formation rate is
6 $10^{-10}$\yrinv, well above $1/t_{hubble} = 1.7 10^{-10}$ \yrinv.
At $z=2$ the typical specific star formation rate is 2 $10^{-9}$ \yrinv,
compared to $1/t_{hubble} = 3 10^{-10}$ \yrinv, off by almost a factor of 10.
This phenomenon has been noted by many authors (e.g., Daddi et al. 2007, 
Dav\' e, 2007).
Taken at face value, it suggests that these
galaxies may have formed the bulk of their stars in a very short time.
We note that some authors have argued that the star formation rates are
systematically
over-estimated by a factor of 2-3, either because too many stars are
produced (Wilkins et al. 2008), or because the specific star formation
rates are much higher than theoretical models (Dav\' e 2007).

\figsubeleven

If we accept the high star formation rates, we have to conclude that the
time scale of star formation is getting very close to the orbital times
of these galaxies. 
We estimate the orbital time simply from the inferred velocity
dispersion and the size. 
We emphasize that they are uncertain at high redshift, 
as the velocity dispersions or
circular velocities are not directly measured.
We show the ratio of the star formation time over
the orbital  time in Fig. 12. As can be seen, at redshifts below 1, the
star formation time is $>30$ times the orbital time, but at $z>1.5$ the
ratio gets close to 3. This is  so short, that it is
unlikely that the gas is settled in a cold disk.
Simulations would obviously be needed to understand better the exact
dynamical state.

\figpreeleven

Furthermore, we can roughly estimate the gas fractions of these galaxies,
by assuming that the gas has the same lengthscale as the (blue) light,
and by assuming the Kennicutt (1998) relation between gas surface density, and
star formation rate per area. 
Even though the estimated gas masses
must be considered to be very uncertain (as the sizes are uncertain, and
the Kennicutt relation has not been established at $z=2$), the results
can be used for at least consistency checks with models which assume
the Kennicutt relation as a general recipe for  star formation.
We note that Bouch\' e et al. (2007) confirmed the Kennicutt relation
for sub-millimeter galaxies at $z=2-3$.
The resulting gas mass to stellar mass ratios for the star forming 
galaxies are shown
in Fig. 12. As can be seen, the gas to star ratios for the star forming
galaxies increase from 0.2 at $z=0$  to 1 at $z=1.5$.
Even if the star formation rates have been overestimated by a factor of 2 
at $z=1.5$, this result will not change much.
As a result, we have to conclude that the general relations between star
formation and gas density at $z=0$ imply that the typical  
star forming galaxies at
$z=1.5$ and above have very significant gas fractions. A similar result
had been found before by Erb et al. (2006), based on H$\alpha$ spectroscopy.

Another issue related to these strongly star forming galaxies is how
long the star formation episodes last. If the star formation rate would
be constant, the specific star formation rate would decline very rapidly
(on a timescale of 1/specific star formation rate). As the 
epoch in which the specific
star formation rate is high would last only briefly, the universe would
be expected to be dominated by galaxies with specific star formation rates
comparable to the inverse of the Hubble time.
This is obviously not what is observed. Either the specific star formation
rates are over-estimated, or the galaxies undergo brief bursts punctuated
by periods of low star formation, or the galaxies form with constant
specific star formation rate.
The first explanation requires an over-estimate by a factor of 3 -
which is entirely possible, given the uncertainties with the conversion
to bolometric luminosities, and uncertainties in the IMF.
The second mechanism would imply that we would have to see a substantial
number of "dead" galaxies, with the same masses and sizes of the 
"live" galaxies. However, one of the basic results of this paper is
that we see a good correlation between the structure and specific star 
formation rate of the galaxies, and this would be washed out if
galaxies were to undergo frequent bursts. The only way to explain this
would be to assume that the sizes and possibly masses of the galaxies
have been estimated wrongly for the star forming galaxies. 
If high redshift galaxies are multi-component, with
high gas fractions, and large dust masses, it might be possible that their
derived quantities from just optical light might be seriously wrong.
Even in the nearby universe this plays a role: color gradients in spirals
are significant, and this does imply that the optical half-light radius
can deviate significantly from the half-mass radius. 
The derived gas surface densities for the $z\approx 2$ 
galaxies imply typical absorption
in the V-band of $A_V \approx 10$, suggesting that a significant amount of
starlight is completely obscured.
This result appears entirely reasonable, but is somewhat inconsistent
with
the result that SED fits to the restframe UV-optical-near-IR give
good estimates of the total star formation rate, suggesting that
the galaxies are semi-transparent (e.g., Daddi et al. 2007).
Obviously, this should investigated further.

Alternatively, the high specific star formation rates are correct,
and many galaxies form very rapidly with
nearly constant, very high specific star formation rate.
This implies rapid expontial growth on a short timescale (e.g.,
Daddi et al. 2007): in this
case, they would increase by a factor of 10 in mass in a time of
2.3 * 1/SSFR, corresponding to 1 Gyr at $z=2$. For the $z=2$ galaxies,
they would be less massive  by a factor of 10 at $z=2.8$.
This does not appear completely unreasonable, but we have to note
that current models of galaxy formation do not predict this type of
behaviour (e.g., Dav\' e  2007).

\subsection{Comparing the evolution of the mass-size relation with models}

As we have seen, the mass-size relation of galaxies evolves smoothly
from $z=0$ to $z=3.5$, like $\re \propto (1+z)^{-0.5}$ at a given mass.
There is  evidence that the relation evolves faster for high
masses,
and quiescent galaxies also tend to evolve faster.
The general evolution is fairly close to what is predicted for 
simple disk galaxy evolution, where $r \propto 1/H(z)^{2/3}$ (Mo et al. 1998).
In the redshift interval from 0 to 3.5, we find approximately
$1/H(z)^{2/3}  \propto 1/(1+z)^{0.79}$, fairly close
to the observed size evolution.
We note that in a detailed analysis, Somerville et al. (2008) found
that
the expect rate of evolution is somewhat slower than $1/H(z)^{2/3}$, 
more consistent with the result obtained here.
The fact that the general trend holds can be taken as evidence that
the simple scaling of dark matter halos also determines the scaling of
the  galaxies.

Hence the question remains open whether the very compact quiescent galaxies
at high redshift have formed in  intrinsically very different ways
compared to their cousins at low redshift.
Kochfar \& Silk (2006) invoked strong dissipation at high redshift to
explain the very small sizes of galaxies at high redshift.
The results obtained here suggests that star forming galaxies
at redshifts as low as $z=1.5$ also had very high gas fractions, 
potentially undercutting this explanation for the evolution.

If we assume that our trends persist to $z=4$, we would be let to conclude
that the compact galaxies at $z=2$ formed their stars around $z=4$:
Their average  specific star formation rate is approximately
$10^{-10}$\yrinv at $z=2$. If we assume that the specific star formation
rates
evolve like $(1+z)^{3.5}$ (as for the specific star formation rate 
at the threshold surface density),  we find that the
specific star formation rate
would be $1/t_{hubble}$ 
when the universe was about 1.6 times smaller, i.e., at $z=3.8$. 
Obviously, this is an  uncertain extrapolation of the relations found at
$z<3$.

The stellar ages of the quiescent galaxies at $z=2.5$ have been
estimated around 1Gyr  (Kriek  et al. 2008), this would imply a
similar formation
redshift of 4.

\subsection {Relation to sub-mm galaxies}

One of the striking results in this work is that the star forming galaxies
at $z>1.5$ are large compared to the quiescent galaxies,
in agreement with earlier studies (e.g., Zirm et al. 2007 and
Toft et al. 2007). Furthermore,
the star forming galaxies have very high star formation rates and 
short specific star formation times, and
would be called starbursts if 
they occured in the nearby universe. 
One of the interesting questions is whether  they
are related to the ULIRGS at low redshift, and the  sub-mm galaxies at 
high redshift.
 We note that ULIRGS in the nearby universe have very concentrated
star formation in very small volumes, with typical sizes significantly
smaller than 1 kpc (Tacconi et al. 2006, and references therein). 
Hence the high redshift star forming
galaxies
have comparable IR luminosities (typically $10^{12}\ L_\odot$), but
are
inferred to have a very 
different structure than the local sub-mm galaxies.
Unfortunately, catalogues of sub-mm sources in the CDF-South are not
yet available.
Size measurements of sub-mm galaxies at high redshift are rare.
Tacconi et al. (2006, 2008) 
found  typical sizes of the gas smaller than 2 kpc at 
$z=2$, which is small for the star forming galaxies found here at the 
same redshift.
A detailed comparison of the sizes measured for the stars, gas, and
star formation region would be valuable.
As the space density of the sub-mm galaxies at high redshift is low,
it may be that they simply lie in the tail of the size distribution 
for star forming galaxies.
Larger samples are needed to test this.

\subsection{Caveats and further work}

The results obtained here illustrate the power of structural studies of
galaxies. It is appropriate to discuss the potential errors that may have
occured in the derivation of the masses, sizes, and star formation rates.

1) masses: it is well known that the  stellar masses derived from
sed fitting are rather uncertain, even though they are the most stable
outcome of such fits. Problems include the uncertainties in stellar population
models themselves (without considering differences in star formation
history,
for example Maraston 2005 versus Bruzual \& Charlot 2003),
uncertain star formation histories, uncertain geometries of the dust,
and potential correlations between absorption and stellar age, and the
possibility that some components are entirely hidden throughout the 
rest-frame near-ir due to very high extinction. In addition, the IMF
of the stars may vary 
(e.g., 
van Dokkum 2008,
Dav\' e 2008, 
Wilkins\& Hopkins et al. 2008).
Only more detailed observations can provide answers. High resolution 
observations with HST can determine whether galaxies have strong color
gradients,  complicating
the SED fitting. Direct spectroscopy is urgently needed to establish the
stellar velocity dispersions of the compact, quiescent galaxies, and
the star forming galaxies (which will be even harder).
High spatial resolution observations of molecular lines can provide detailed
information on the mass distribution in the inner parts, and can maybe
provide insight into hidden populations, whether old or young.
Rest-frame optical spectroscopy can potentially probe the mass distributions in
the outer parts.

2) Sizes: the sizes used here are sizes measured in the rest-frame optical.
For the small galaxies, the sizes are just barely resolved.
Obviously, higher resolution imaging is required for the rest-frame optical
sizes at $z>1.5$, and, in addition, the determination of color gradients
is important to see whether the measured sizes could be affected significantly
by color gradients. The color gradients in the nearby universe vary from
0.07 in B-R per dex radius for ellipticals to 0.2 in g-r per dex radius for 
late-type spirals, (e.g., Franx \& Illingworth 1990, 
Peletier et al. 1990, 
de Jong 1996). 
If we use simple relations between mass-to-light ratio and color, we find
that the half mass radii are smaller by
about a factor of 0.87 and 0.56, respectively. This is insufficient to wash out
the effects which we have seen, but larger effects could be present at
high redshift.  Simulations can also play a role here: Hopkins et al. (2008)
have shown that the optical sizes of the merger remnants can be smaller
by a factor of 2 than the true half-mass size. This is caused by
the concentrated young population of stars formed at the end of the merger.
Joung et al. (2008) find that the
apparent
sizes of simulated star forming galaxies at $z=3$ are a factor of 3
higher
due to extinction by dust.
Obviously, the  interpretation of the apparent
sizes may not be straightforward, and other diagnostics may be needed
(e.g. the spatial distribution of the star forming regions).
The models by Guo \& White (2008) predict the correct qualitative rise
in
specific star formation rate, but  unfortunately do not predict sizes.

3) Star formation rates:
It is well known that the derivation of reliable star formation rates is
still very hard. The extrapolation of the measured 24 $\mu$m flux to
a total bolometric 	IR flux  is uncertain (although the results of
Papovich et al. 2007 imply that a simple linear relation may suffice). 
Studies with Herschel may improve upon this situation.
Furthermore, even if the
bolometric flux is well determined, the star formation rate is not, as
the IMF may vary with redshift. For example, the IMF may vary at very
high masses (around the masses of O stars, 50 $\Msun$,), at the full mass
range between 1 and 50 $\Msun$, and at the low mass range (as suggested
by Dav\' e 2008 and van Dokkum 2008).
There is direct evidence that the star formation rates estimated traditionally
are too high: the mass in stars seems to be over-produced (Wilkins et al
2008, Dav\' e 2008). 
van Dokkum (2008) 
emphasized that
a change in the IMF results in both changes in the derived star formation
rate, and the derived masses, and the changes depend on the exact form
of the IMF evolution. Omitting or adding low mass stars ( $m < 1 \Msun$) 
essentially does not do very much, as all masses and star formation rates
are changed by the same factor, 
and the specific star formation rates remain the same
(and too high). Changes in the characteristic mass of Chabrier  type
IMFs will change the star formation rates more than the masses, which
is the type of change that is desired; and changing the slope above $m= 1 \Msun$
will have the largest effect on the derived star formation rates, and smaller
effects on the masses. Obviously, the exact investigation of these effects
is beyond the scope of this paper. 
The calibration of the stellar masses of quiescent
galaxies at high redshift can play an important role in these investigations,
in addition to the high resolution dynamical studies of gas and stars to
decompose galaxies.
The comparison of many different star formation indicators will also provide
further insight in this issue.

4) cosmic variance: the current field is fairly small, and does not
have
large numbers of high mass galaxies. It will obviously be important
to study the same relations on larger fields, thereby making a
fairer sample of the universe. Furthermore, such studies will allow
the
determination of the distribution of surface densities, and inferred velocity
dispersions as a function of redshift.

\section{Implications for high redshift galaxy studies: biases and consequences}

\figpretwelve
\figsubtwelve

The correlations which we found above have important consequences for
observational studies. First of all, many studies have flux limits
or color selection criteria which may make them to pick up specific
subsamples in the space of  mass, size, and specific star formation
rate.
For example, studies have now begun of the H$\alpha$ emission line
kinematics and spatial distribution (e.g., Erb et al. 2003, 2006,
Genzel et al. 2006, F\"orster Schreiber
et al. 2006, Kriek et al. 2006, Law et al. 2007, Wright et al. 2007).
Many of these studies impose a flux limit on the H$\alpha$ emission
line flux before the detailed observations are performed.
The H$\alpha$ fluxes have not been measured for this sample, but we
can estimate them using the star formation rate, and the estimated
extinction $A_V$. We used the conversion by Kennicutt (1998) to
transfer the star formation rate to unobscured H$\alpha$ flux.
We impose a flux limit of $10^{42}$ ergs sec$^{-1}$,
and we show the selected galaxies in the left hand panel of Fig. 13.
As is clear, the galaxies with strong H$\alpha$ 
are preferentially large, and have high
specific star formation rates for galaxies at that mass, as might be
expected. Their star formation timescales are generally a few times
the dynamical time, and hence their dynamical state is not typical for
the median galaxy at that redshift.

Another, often used selection technique is the Lyman Break selection
technique, or BM-BX selection employed by Steidel et al. (1999, 2004),
and many other authors. We show in the right hand panel those galaxies
which satisfy the criterion $B-R < 1.2$, and $R < 25.5$, which are
the criteria often used for these studies. Again, it is clear that the
selection picks up preferentially large  galaxies from the samples.
Near-IR spectroscopic follow-up of these samples introduces
an additional bias (through the flux limit on H$\alpha$).

Apart from the selection effects discussed above, the results imply
that the interpretation of the SEDs of the galaxies may be
over-simplified
in the current models. As the relation between mass and radius
evolves with time, and as galaxy masses increase with time, it is
likely that galaxies have old subcomponents which are smaller than
the new additions. If so, the subcomponents will have different 
extinction as well, and the modeling of the galaxies as simple
populations
with constant extinction may give biased results.
Without more detailed information, it is hard to quantify these
effects,
and first of all, direct dynamical mass estimates are needed.
However, simulations can also be used to estimate the sizes of such
effects. We notice that Wuyts (2007a) found that the stellar mass can be
underestimated by a factor of 2 in the phase
of strong star formation in  a gas rich merger. 
These effects are in addition to the known
uncertainties in stellar populations models.

\section{Summary and conclusions}

We have shown that massive galaxies from $z=0$ out to $z=3.5$
have a strong correlation between size and color and size and specific
star formation rate, at a
given mass. Galaxies with high specific star formation rates are
large, galaxies with low specific star formation rates are small. 
At increasing redshifts, the overall specific star formation rates go up.

In general, the specific star formation rates correlate better 
with surface density, 
and inferred velocity dispersion, than with  mass. This suggest that surface density, or
inferred velocity dispersion, is the driving parameter. 
We find that, as expected, specific star formation rates  at a
given surface density increase with redshift.
We identified a threshold surface density at each redshift interval: 
below the threshold the specific star formation rates are high with
little
variation, above the threshold density galaxies
have low specific star formation rates.
As expected, the threshold increases with redshift: high specific
star formation occurs at higher and higher surface density with increasing
redshift.
As a result, we find that many  galaxies which are on the red sequence
at $z=0$ are star forming at $z=1$.

Furthermore, the sizes of galaxies at a given mass decrease with redshift
steadily from $z=0$ to $z=3$.  This overall evolution shows that
galaxies grow inside out. As this growth also occurs for 
'red and dead galaxies', it suggests that these galaxies keep evolving - and
are never just 'passively' evolving. 
In short, all galaxies show 'up-sizing', and the most massive galaxies
show the strongest evidence for it.
There is no evidence in this sample that very massive galaxies do not
evolve
between $z=1$ and $z=0$ (Scarlata et al. 2007), but obviously larger
area studies can address this issue better.
The very small, red galaxies at $z=2-3$ 
have to evolve into larger galaxies by $z=0$ through merging,
accretion,
and star formation.
Their small sizes may simply be due to the fact that they are on the tail
of the distribution at $z=2-3$, where all sizes are smaller.
Two processes are likely  responsible for their small sizes:
the halos were smaller and denser, and additionally their gas fractions
were higher, and dissipation during
merging was stronger, as suggested by Khochfar \& Silk (2006). 
The observed evolution  is also consistent with merger simulations (e.g.,
Hopkins et al. 2008).

The similarity in structural relations for galaxies from $z=0$
to $z=2.5$, and the
fact that galaxies grow inside out suggests that one form or another of
the Hubble sequence persists to $z=2.5$, and possibly beyond. The older
stars likely dominate in the centers, and younger stars are 
likely distributed over
a larger radius - similar to bulges and disks in spiral galaxies.
Although we don't have the resolution to establish this directly for
the galaxies, the evolution of the mass-size relation with redshift
also strongly supports such inside-out growth of galaxies.
The exact dynamical state of high redshift galaxies still needs to
be determined. 

The multicomponent nature of galaxies suggest that modeling these galaxies
is much harder, as the different populations have different spatial 
distributions, and therefore different extinction, star formation
history, etc.
Analysis of simulations suggests that this can lead to under-estimates of
the masses through SED fits, and over-estimates of the sizes (e.g.,
Wuyts et al. 2007a).

The galaxies with very high star formation rates at $z \ge 1.5$ likely
have very high ratios  of gas mass to stellar mass (approximately 1 or
above).
Furthermore, they are large ($\approx 3 kpc$), and therefore different
from  ULIRGs in the nearby universe, which have typical sizes of
the star formation regions of $<< 1kpc$.
However, these high redshift galaxies may not be simple cold disks:
their star formation time scale is only a few dynamical times, and
therefore the gas had barely time to settle in discs, if accretion is
causing the very high star formation.
Alternatively, the star formation may have been overestimated, and
this would allow the gas more time to settle.

Obviously, these results call for many follow-up observations. First and 
foremost, the mass estimates must be improved, hopefully through dynamical
mass estimators, either through near-ir spectroscopy, or spectroscopy
with ALMA.
Second, higher resolution Near-IR imaging can determine 
the structure of nature of the high redshift galaxies better.
High resolution imaging with ALMA will be able to establish the distribution
of the star formation across the galaxies, and spectroscopy will
allow the determination of the gas content and gas distribution.

Third, it will be important to extend this work to even higher redshifts,
where samples selected in the rest-frame optical are very rare, and 
structural analyses absent.  Fourth, the environment of the high
redshift  galaxies
needs to be determined. This will allow a study of the
the relation between structure, 
star formation history and environment  at high
redshift. At low redshifts, environment plays an important role in
setting the star formation rate (e.g., Kauffmann et al. 2004), and
a full understanding requires an extension to high redshift.
Finally, a determination of evolution of the surface density function 
can help to
confirm the simple picture in which the centers of massive galaxies
formed first.

The results show that stellar surface density, or inferred velocity 
dispersion, is
one of the main driving parameters of galaxy evolution.
Studies of the correlation of other galaxy properties with these parameters
would be extremely valuable: metallicities, observed circular velocities,
but also correlation length, and environment. Such studies require 
surveys of much larger areas, and extensive spectroscopy.
The velocity dispersions used here are estimated from $M/\re$, and it
remains to be verified whether the two are well correlated at all
redshifts, for all galaxies. If not, it is crucial to determine which
is the driving parameter.



\acknowledgments

The comments of the referee helped to improve the paper.
We thank the Leids Kerkhoven Bosscha foundation for providing travel
support. We thank the Lorentz Center for hosting workshops during
which this paper was written.
We thank  Joop Schaye, Phil Hopkins, Lars Hernquist, Rachel
Somerville for discussions.
Support from NASA grant HST-GO-10808.01-A is gratefully acknowledged.
S. Wuyts acknowledges support from the
W. M. Keck Foundation.







 \appendix

 \section{Appendix A}
\figsubAone

Here we analyze in more detail whether some of our evolutionary
effects may be caused by selection effects.  Trujillo et al. (2006a)
presented the impact of surface brightness selection effects on the
distribution of effective radii for a comparable set of galaxies in
the field of MS1054-03, and found no strong effects.
As our own dataset is very comparable in depth and uses the same
instrument, we don't expect strong selection effects here either.

We have to note, however, that selection effects in the mass-size and
surface density specific star formation rate  planes may be more
complex, as these quantities are derived indirectly from the
observables. To determine whether our evolutionary effects may be
driven by selection biases, we perform an analysis where we use
low redshift samples as a  reference sample, and transpose  the
galaxies
to higher redshift, while keeping their intrinsic properties the same.
This allows us to determine directly whether the evolution observed
here is caused by simple selection effects.

\figpreAone

\subsection {Moving galaxies from $z \approx 1$ to $z \approx2$ }

We first take the galaxies in the redshift bin $0.5 < z < 1.5$ as our reference
sample. We increase the redshift of each galaxies by 1 unit in redshift,
while maintaining  constant intrinsic absolute magnitudes, and
apparent size. For each galaxy, we then re-determine whether it has
sufficient signal-to-noise ratio and $K$ band flux
to be included in our sample.
The resulting distributions in the mass-size plane, and the surface
density-specific star formation plane are shown in the top row of Fig. A14.
The open squares indicate galaxies which are still detected at $z
\approx 2$, the small symbols are galaxies which are lost in the process.
It is clear that in this  simulation the incompleteness is very strong
below a mass of 3\ $10^{10}$  $\Msun$. Even above that mass significant
numbers of galaxies are missing. However, it is also obvious from the
plot that there is no bias towards losing large galaxies.  We
quantified this by determining the median radii of all
galaxies,
and the galaxies still selected when shifted at $z\approx 2$. This two
radii are the same within 2 \%.

The upper right figure shows the distribution in the plane of specific
star formation against surface density. Again, it is clear that the
galaxies which persist when shifted are not strongly biased.

\subsection {Moving galaxies from $z\approx2$ to $z\approx3$ }

In a similar analysis, we shifted the galaxies at $z\approx 2$ to
$z\approx 3$. The result is shown in the bottom row of Fig. A14.
Again, a large fraction of galaxies is lost. This should not come as a 
surprise, as the simulation assumes no evolution, and therefore
may very well overpredict the fraction of lost galaxies.
We note a small bias towards keeping large galaxies in the sample -
exactly
the opposite from what might be expected from surface brightness
selection
effects. This is caused by the fact that red galaxies are lost the
fastest,
as they are faintest in $K$ (and all other bands), for the same mass.
Hence the median radius of the galaxies still detected after shifting
them
is slightly higher (by 15\%), than the median radius of all galaxies
more
massive than 6\ $10^{10}$ $\Msun$.
Overall, this is a small effect, and we ignore it in the analysis.
We also see that the distribution in the plane of specific star
formation
against surface density is not affected by these selection  effects.

In short, we conclude that the selection effects are not likely to
cause
the strong evolution in specific star formation rate and size.
This is likely caused by the fact that the point spread function is
rather large, compared to the size of the galaxies, and hence 
strong surface brightness selection effects are not likely to play a
role.
It is also clear that the main selection effect is caused by the $K$
band limiting depth. This causes quiescent galaxies of a given mass to
be missed earlier than strongly star forming galaxies. As these
quiescent galaxies are small, it causes us to miss small galaxies
first below the mass limit at which we are complete.

\vfill\eject
\vbox{\hsize=16truecm
{ TABLE 1: LIMITING MAGNITUDES FOR THE GOODS-SOUTH IMAGING}
\vskip 0.7truecm
\halign { \hskip 3truecm # \hfil & # \hfil & # \hfil \cr
Camera & Filter &  Magnitude$^a$\cr
\cr
WFI	& $U_{38}$		& 26.06 \cr
WFI	& $B$			& 27.21 \cr
WFI	& $V$			& 26.88 \cr
WFI	& $R$			& 26.99 \cr
WFI	& $I$			& 25.00 \cr
ACS	& $F435$		& 27.29 \cr
ACS	& $F606$		& 27.42 \cr
ACS	& $F775$		& 26.87 \cr
ACS	& $F850LP$		& 26.51 \cr
ISAAC	& $J$			& 25.43 - 26.03 \cr
ISAAC	& $H$			& 25.00 - 25.55 \cr
ISAAC	& $K_s$			& 24.63 - 25.57 \cr
IRAC	& ch1			& 26.15 \cr
IRAC	& ch2			& 25.66 \cr
IRAC	& ch3			& 23.79 \cr
IRAC	& ch4			& 23.70 \cr
MIPS	& 24 $\mu$m	 	& 21.30\cr}
}
$^a$ The limiting magnitude is the total AB magnitude for
point sources detected at 3-$\sigma$.

\end{document}